\documentclass[useAMS,usenat]{aa}
\usepackage{txfonts}
\usepackage{natbib}
\bibpunct{(}{)}{;}{a}{}{,}
\usepackage{longtable}
\usepackage[dvips]{graphicx}

\begin{document}
\title{Luminosity functions of LMXBs in different stellar environments}
\author{Zhongli Zhang \inst{1} \and Marat Gilfanov \inst{1,2} \and Rasmus Voss \inst{3} \and Gregory R. Sivakoff
\inst{4} \and Ralph P. Kraft \inst{5} \and Nicola J. Brassington \inst{5,6} \and Arunav Kundu \inst{7} \and Andr{\'e}s Jord{\'a}n \inst{8,5} 
\and Craig Sarazin \inst{4}}
\institute{Max-Planck Institut f\"{u}r Astrophysik, Karl-Schwarzschild-Stra\ss e 1, D-85740 Garching, Germany 
\and Space Research Institute, Russian Academy of Sciences, Profsuyuznaya 84/32, 117997 Moscow, Russia 
\and Department of Astrophysics/IMAPP, Radboud University Nijmegen, PO Box 9010, NL-6500 GL Nijmegen, the Netherlands
\and Department of Astronomy, University of Virginia, P.O. Box 400325, Charlottesville, VA 22904-4325, USA 
\and Harvard/Smithsonian Center for Astrophysics, 60 Garden Street, MS-67, Cambridge, MA 02138, USA 
\and School of Physics, Astronomy and Mathematics, University of Hertfordshire, College Lane, Hatfield AL10 9AB, UK
\and Eureka Scientific, 2452 Delmer Street Suite 100, Oakland, CA 94602-3017, USA 
\and Departamento de Astronom{\'i}a y Astrof{\'i}sica, Pontificia Universidad Cat{\'o}lica de Chile, Vicu\~na Mackenna 4860, 7820436 Macul, Santiago, Chile 
\\
email:[zzhang;gilfanov]@mpa-garching.mpg.de}
\titlerunning{LMXB LFs in different environments}
\date{Received ... / Accepted ...}

\abstract{} {
Based on the archival data from the $Chandra$ observations of nearby galaxies, we study different sub populations of low-mass X-ray binaries (LMXBs) -- dynamically formed systems in globular clusters (GCs) and in the nucleus of M31 and (presumably primordial) X-ray  binaries in the fields of galaxies. Our aim is to produce accurate luminosity distributions of X-ray binaries in different environments, suitable for quantitative comparison with each other and with the output of population synthesis calculations.
} 
{Our sample includes seven nearby galaxies (M31, Maffei 1, Centaurus A, M81, NGC 3379, NGC 4697, and NGC 4278) and the Milky Way, which  together provide relatively uniform coverage down to the luminosity limit of $10^{35}$ erg/s. In total we have detected 185 LMXBs associated with GCs, 35 X-ray sources in the nucleus of M31, and 998 field sources of which $\sim 365$ are expected to be background AGN. We combine these data, taking special care to accurately account for X-ray and optical incompleteness corrections and the removal of the contamination from the cosmic X-ray background sources, to produce luminosity distributions of X-ray binaries in different environments to far greater accuracy than has been obtained previously.} 
{We found that luminosity distributions of GC and field LMXBs differ throughout the entire luminosity range, the fraction of faint ($\log(L_X)<37$) sources among the former being $\sim 4$ times less than in the field population.  
The X-ray luminosity function (XLF) of sources in the nucleus of M31 is similar to that of GC sources at the faint end but differs at the bright end, with the M31 nucleus hosting significantly fewer bright sources. We discuss the possible origin and potential implications of these results.
} 
{} 

\maketitle



\begin{table*}
\begin{center}
\caption{The sample of external galaxies observed by $Chandra$}
\label{tab:sample}
\renewcommand{\arraystretch}{1.1}
\begin{tabular}{llcccccrccc}

\hline
\hline
Galaxy    &Type & Distance& $N_{\rm H}$ & Study Field & $M_{*}/L_{\rm K}$ & $M_{*}$ & Exp & Sensitivity & dx/dy & Conversion Factor \\
 & &(Mpc)& ($10^{20}$ cm$^{-2}$) & &($M_{\odot}/L_{\rm K,\odot}$) & (10$^{10}M_{\odot}$) & (ks) & (erg s$^{-1}$) & (pixel) & (erg cm$^{-2}$ count$^{-1}$)\\
  (1)    &(2)  &(3)    &(4)   & (5)                &(6)    &(7)   &(8)        &(9)                 &(10)         &(11) \\  
\hline
M31      & Sab &  0.78$\pm0.03$  & 6.7  & $r=11^{\prime}$	   & 0.56  &  2.7 & $\sim$300 & 4$\times$10$^{34}$ & +0.02/-0.37 & $3.4\times10^{-9}$ \\
         &--&--&--& $15^{\prime},9^{\prime},-65^{\circ}$   & --    & 0.60 & $\sim$150 & 7$\times$10$^{34}$ & +0.11/+0.51 &  \\
Cen A    & S0  &  3.4$\pm0.4$  & 8.6  & $r=10^{\prime}$    & 0.76  &  6.4 & $\sim$800 & 6$\times$10$^{35}$ & --   /--    &  $3.5\times10^{-9}$ \\
M81      & Sab &  3.63$\pm0.34$  & 4.2  & $10^{\prime},5^{\prime},-20^{\circ}$ & 0.70  &  6.0 & $\sim$240 & 7$\times$10$^{35}$ & +0.10/+0.49 &  $3.2\times10^{-9}$ \\
Maffei 1 & S0  &  3.0$\pm0.3$  & 85.1 & D25 \& HST         & 0.73  &  1.1 & $\sim$55  & 3$\times$10$^{36}$ & +0.01/+0.29 &  $7.7\times10^{-9}$ \\
N3379    & E1  &  11.1  & 2.8  & D25 \& HST         & 0.83  &  5.4 & $\sim$330 & 4$\times$10$^{36}$ & --   /--    &  $3.1\times10^{-9}$ \\
N4697    & E6  &  11.8  & 2.1  & D25	           & 0.77  & 5.8 & $\sim$200 & 5$\times$10$^{36}$ & --   /--    &  $3.0\times10^{-9}$ \\	  
N4278    & E1-2&  16.1  & 1.8  & D25 \& HST         & 0.79  &  4.2 & $\sim$480 & 6$\times$10$^{36}$ & --   /--    &  $3.0\times10^{-9}$ \\			       
\hline
\end{tabular}
\end{center}
 (1) -- Galaxy name. For M31, the first line is for the bulge region, the second line is for the region in the disk. 
(2) -- Galaxy Type. 
(3) -- Distance  and its uncertainty (when available). References and methods are: M31 -- luminosity function of red clump stars \citep{Stanek1998}; Centaurus A -- Cepheids \citep{Ferrarese2007}; M81 -- Cepheids \citep{Freedman1994}; Maffei 1 -- galaxy fundamental plane \citep{Fingerhut2003}; NGC 3379 -- luminosity function of GCs \citep{Kundu2001}; NGC 4697, NGC 4278 -- surface brightness fluctuation \citep{Tonry2001}.
(4) -- Galactic column density \citep{Dickey1990}.
(5) -- The region used to study XLFs. When three numbers are given, they refer to major, minor axis and position angle. 
(6) -- {\it K}-band mass-to-light ratios derived from \citet{Bell2001}, with $B-V$ colors from the RC3 catalog
\citep{Dev1991} except for Maffei 1, which is from \citet{Buta1983}.
(7) -- Stellar mass in the study field, as calculated from the {\it K}-band magnitudes derived from  2MASS Large Galaxy Atlas \citep{Jarrett2003}. For M31 we used the IRAC/Spitzer data, and the 3.6 $\mu$m flux was converted to  {\it K}-band following \citet{Bogdan2010}. 
(8) -- The total exposure time of $Chandra$ observations. 
(9) -- Point source detection sensitivity estimated from  the incompleteness functions in Fig. \ref{fig:icf}. 
(10) -- Attitude correction.
(11) -- Conversion factor of $Chandra$ count rate to unabsorbed X-ray flux in the 0.5-8 keV band.

\end{table*}

\section{Introduction}
\label{sec:introduction}

It has long been known that there are many more low-mass X-ray binaries (LMXBs\footnote{Throughout this paper we refer to objects that have been actively accreting in recent times (i.e. $\log(L_X)\ga$ 35) as X-ray binaries.}) per unit stellar mass in Galactic globular clusters (GCs) than in the field \citep{Clark1975}.
This fact is conventionally explained as a result of dynamical formation of LMXBs in the high stellar density environment of GCs where the probability of two-body interactions, which scales as $\rho^2_\ast$, is high  \citep{Fabian1975}. 
In the $Chandra$ era this picture received further support from the high specific frequency of LMXBs in GCs observed in nearby external galaxies  \citep[e.g.,][]{Angelini2001,Sarazin2003,Minniti2004,Jordan2007b}. Also, a significant ``surplus" of LMXBs was detected 
in the nucleus of M31, with the spatial distribution of compact X-ray sources following the ``$\rho^2_\ast$" law \citep{Voss2007a}.
The stellar density is low outside of GCs and the nuclear region of galaxies, with a correspondingly lower probability of stellar interaction, therefore primordial formation is thought to be the main formation process for LMXBs in the main bodies of galaxies. Their 
volume densities follow the distribution of stellar mass \citep{Gilfanov2004}. 

Although the above picture is attractive in its simplicity, there is a plausible alternative scenario: The entire population of LMXBs in galaxies, including those in the field may have been produced dynamically in GCs and later expelled into the field. Although the debate is 
still going on \citep{white2002,Kundu2002,Kundu2007,Irwin2005,Juett2005,Humphrey2008}, several strong arguments have been presented which suggest a (significant) fraction of field LMXBs formed in situ via primordial binary formation. These include the difference in spatial distributions of field LMXBs and GCs \citep[e.g.][]{Kundu2007} and the lack of correlation between the specific frequency of field LMXBs and that of GCs \citep[e.g.][]{Juett2005}. At the same time, \citet{Humphrey2008} came to the opposite conclusion. 
 The recently found evidence that the X-ray luminosity function (XLF) of GC-LMXBs may differ from that of field LMXBs  \citep{Voss2007a,Voss2009,Woodley2008,Kim2009} adds to this debate. Although some caveats are in order, differences in the luminosity distributions of the GC and field binaries suggest that the two sub populations of LMXBs may have different formation and/or evolution histories \citep{Voss2009}.

Differences in the luminosity distributions of LMXBs in GCs and in the field may be most obvious in the low-luminosity
($\log(L_X)\la 37$) domain. Thus the reliable detection and quantitative study of any possible differences in the XLF demands special care in the treatment of incompleteness effects and the removal of the cosmic X-ray background sources (CXB). Another difficulty, of a more fundamental nature, is the statistical noise caused by the small numbers of sources.  Although the majority of previous investigations seem to converge in their conclusions, with a few exceptions \citep[e.g.,][]{Voss2009} most of these studies have marginal statistical significance. However it is difficult to achieve higher quality statistics by studying individual galaxies that host a limited number of sources. Massive ellipticals with their large GC populations could avoid this difficulty. But the long distances to the best candidates require deep X-ray observations, in the Msec range, to reach the required depth. Such data sets are not available in the $Chandra$ archive.
However, the potential impact of accurately determining the luminosity distributions of X-ray binaries located in different environments on our understanding of the formation and evolution of LMXBs is high. This motivated us to attempt to produce the most accurate LF of GC-LMXBs to date by combining $Chandra$ data for multiple galaxies. 
To this end we undertook a systematic survey of nearby galaxies with sufficient numbers of LMXBs and GCs. The results of this study are reported below.
The paper is structured as follows. We describe our selection criteria and resulting sample in section \ref{sec:sample} and the data preparation and analysis in section \ref{sec:analysis}. In section \ref{sec:combined_xlf} we describe the procedure used to combine XLFs. Results are presented and analyzed in section \ref{sec:result} and discussed in 
section \ref{sec:discussion}.  Section \ref{sec:conclusion} lists the conclusions.

\begin{table*}
\begin{center}
\caption{The list of $Chandra$ observations analyzed in this paper. } 
\label{tab:observation}
\begin{tabular}{lrrr|lrrr|lrrr}
\hline
\hline
Galaxy & Obs-ID & Instrument & Exp. (ks) & Galaxy & Obs-ID & Instrument & Exp. (ks) & Galaxy & Obs-ID & Instrument & Exp. (ks) \\
\hline
M31(1) & 0303 & ACIS-I & 12.01      & M31(1) & 7064 & ACIS-I & 29.07 & M81     & 5937 & ACIS-S & 12.16  \\
M31(1) & 0305 & ACIS-I & 4.18       & M31(1) & 7068 & ACIS-I & 9.62  & M81     & 5938 & ACIS-S & 11.96  \\
M31(1) & 0306 & ACIS-I & 4.18       & M31(1) & 7136 & ACIS-I & 4.96  & M81     & 5939 & ACIS-S & 11.96  \\
M31(1) & 0307 & ACIS-I & 4.17       & M31(1) & 7137 & ACIS-I & 4.91  & M81     & 5940 & ACIS-S & 12.13  \\
M31(1) & 0308 & ACIS-I & 4.06       & M31(1) & 7138 & ACIS-I & 5.11  & M81     & 5941 & ACIS-S & 11.96  \\
M31(1) & 0309 & ACIS-S & 5.16       & M31(1) & 7139 & ACIS-I & 4.96  & M81     & 5942 & ACIS-S & 12.11  \\
M31(1) & 0310 & ACIS-S & 5.14       & M31(1) & 7140 & ACIS-I & 5.12  & M81     & 5943 & ACIS-S & 12.17  \\
M31(1) & 0311 & ACIS-I & 4.96       & M31(1) & 8183 & ACIS-I & 4.95  & M81     & 5944 & ACIS-S & 11.96  \\
M31(1) & 0312 & ACIS-I & 4.73       & M31(1) & 8184 & ACIS-I & 5.18  & M81     & 5945 & ACIS-S & 11.72  \\
M31(1) & $^*$1575 & ACIS-S & 38.15  & M31(1) & 8185 & ACIS-I & 4.95  & M81     & 5946 & ACIS-S & 12.17  \\
M31(1) & 1577 & ACIS-I & 4.98       & M31(1) & 8191 & ACIS-I & 4.95  & M81     & 5947 & ACIS-S & 10.84  \\
M31(1) & 1581 & ACIS-I & 4.46       & M31(1) & 8192 & ACIS-I & 5.09  & M81     & 5948 & ACIS-S & 12.18  \\
M31(1) & 1582 & ACIS-I & 4.36       & M31(1) & 8193 & ACIS-I & 5.16  & M81     & 5949 & ACIS-S & 12.18  \\
M31(1) & 1583 & ACIS-I & 5.00       & M31(1) & 8194 & ACIS-I & 5.04  & M81     & 9122 & ACIS-S & 10.04  \\
M31(1) & 1585 & ACIS-I & 4.95       & M31(1) & 8195 & ACIS-I & 4.95  & Maffei 1& 5619 & ACIS-S & 55.75  \\
M31(1) & 1854 & ACIS-S & 4.75       & M31(2) & 0313 & ACIS-S & 6.05  & N3379   & 1587 & ACIS-S & 31.92 \\
M31(1) & 2895 & ACIS-I & 4.94       & M31(2) & 0314 & ACIS-S & 5.15  & N3379   & $^*$7073 & ACIS-S & 85.18  \\
M31(1) & 2896 & ACIS-I & 4.97       & M31(2) & 1576 & ACIS-I & 4.95  & N3379   & 7074 & ACIS-S & 69.95 \\
M31(1) & 2897 & ACIS-I & 4.97       & M31(2) & 1580 & ACIS-S & 5.13  & N3379   & 7075 & ACIS-S & 84.18 \\
M31(1) & 2898 & ACIS-I & 4.96       & M31(2) & 1584 & ACIS-I & 4.97  & N3379   & 7076 & ACIS-S & 70.14 \\
M31(1) & 4360 & ACIS-I & 4.97       & M31(2) & 2049 & ACIS-S & 14.76 & N4697   & 784  & ACIS-S & 39.76 \\
M31(1) & 4678 & ACIS-I & 4.87       & M31(2) & 2050 & ACIS-S & 13.21 & N4697   & $^*$4727 & ACIS-S & 40.45 \\
M31(1) & 4679 & ACIS-I & 4.77       & M31(2) & 2051 & ACIS-S & 13.80 & N4697   & 4728 & ACIS-S & 36.16 \\ 
M31(1) & 4680 & ACIS-I & 5.24       & M31(2) & 2894 & ACIS-I & 4.72  & N4697   & 4729 & ACIS-S & 38.61 \\
M31(1) & 4681 & ACIS-I & 5.13       & M31(2) & 2899 & ACIS-I & 4.97  & N4697   & 4730 & ACIS-S & 40.58 \\
M31(1) & 4682 & ACIS-I & 4.93       & M31(2) & 2901 & ACIS-I & 4.68  & N4278   & 4741 & ACIS-S & 37.94 \\ 
M31(1) & 4719 & ACIS-I & 5.10       & M31(2) & 2902 & ACIS-I & 4.76  & N4278   & $^*$7077 & ACIS-S & 111.72 \\
M31(1) & 4720 & ACIS-I & 5.14       & M31(2) & $^*$4536 & ACIS-S & 54.94 & N4278   & 7078 & ACIS-S & 52.09 \\
M31(1) & 4721 & ACIS-I & 5.16       & M81    & $^*$0735 & ACIS-S & 50.56 & N4278   & 7079 & ACIS-S & 106.42 \\
M31(1) & 4722 & ACIS-I & 4.87       & M81    & 5935 & ACIS-S & 11.12 & N4278   & 7080 & ACIS-S & 56.54 \\
M31(1) & 4723 & ACIS-I & 5.05       & M81    & 5936 & ACIS-S & 11.55 & N4278   & 7081 & ACIS-S & 112.14 \\

\hline
\end{tabular}
\end{center}
M31(1) is the bulge region and M31(2) is the disk region. The observations marked by "*" were used as the reference when combining the data.
\end{table*}

\begin{figure*}
\begin{center}
\resizebox{0.33\hsize}{!}{\includegraphics[angle=0]{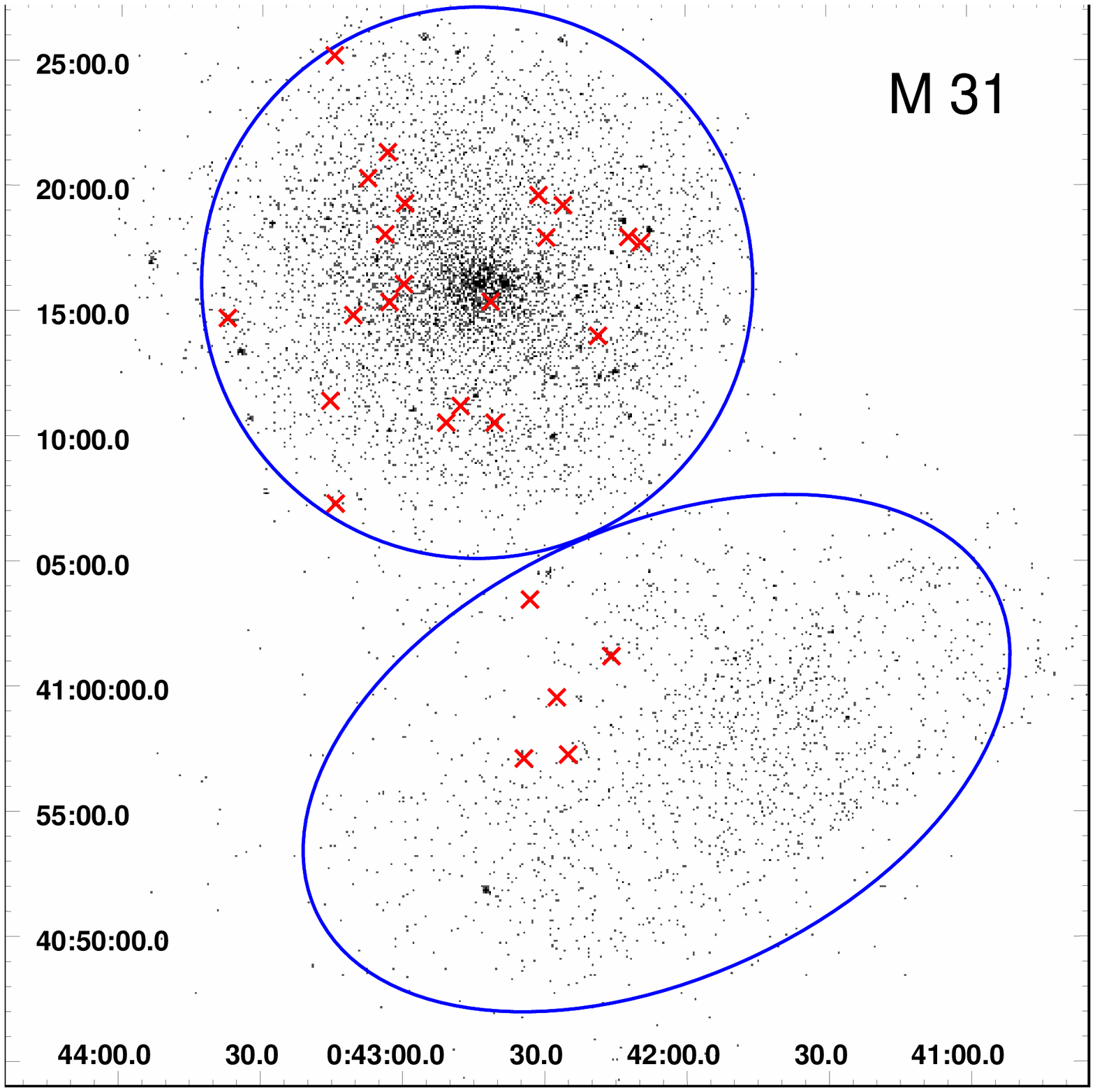}}
\resizebox{0.33\hsize}{!}{\includegraphics[angle=0]{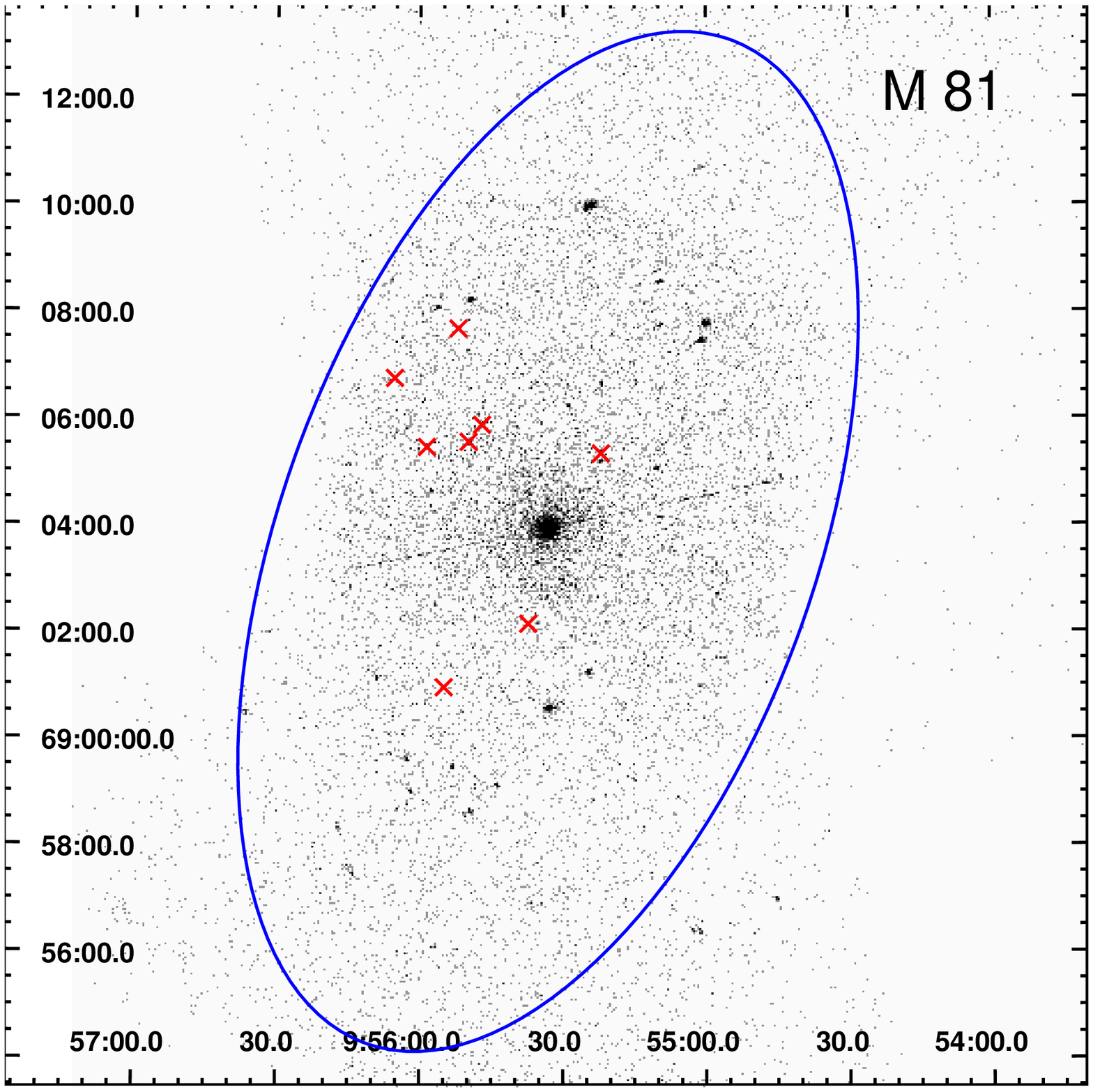}}
\resizebox{0.33\hsize}{!}{\includegraphics[angle=0]{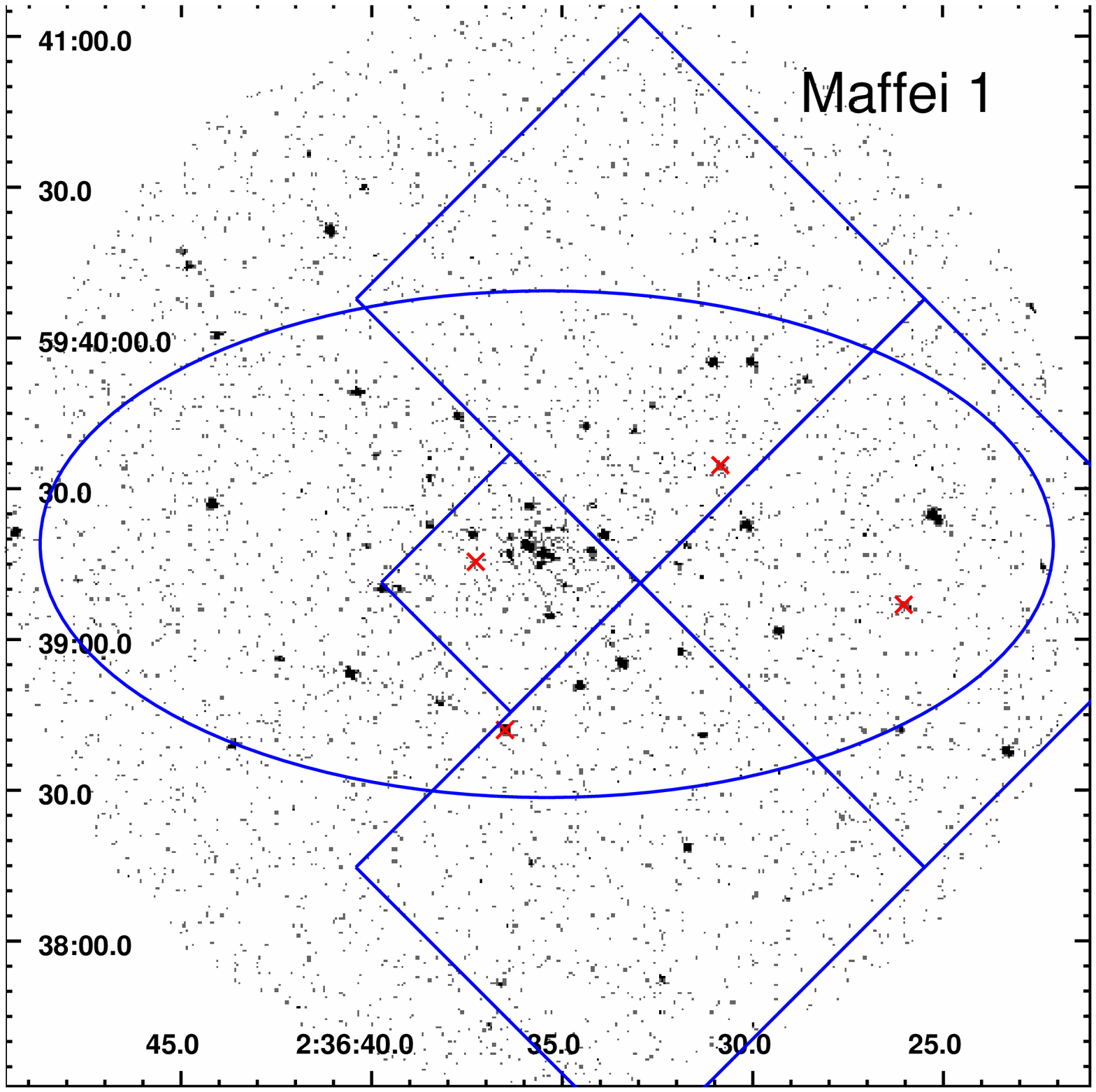}}\\
\resizebox{0.33\hsize}{!}{\includegraphics[angle=0]{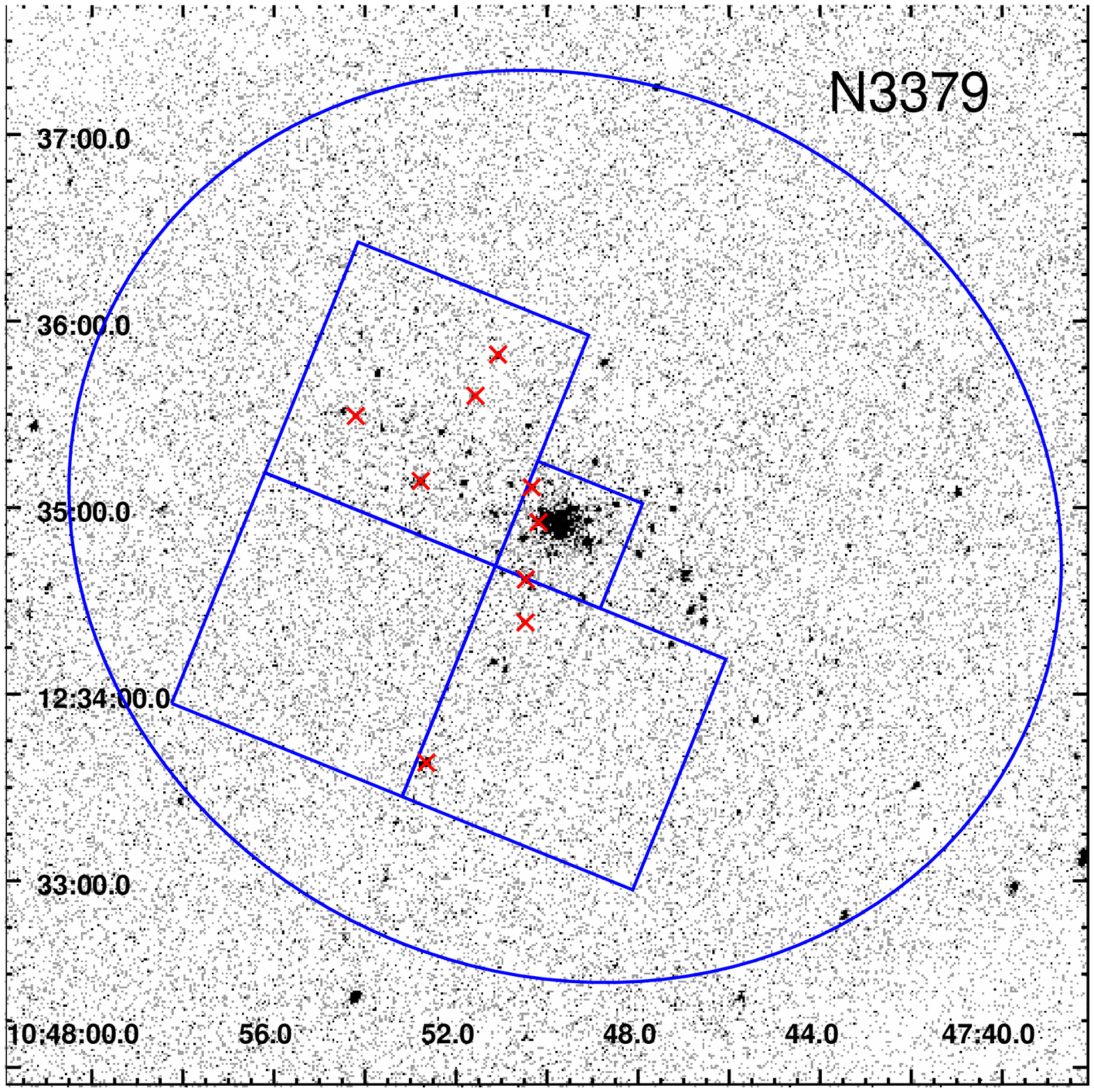}}
\resizebox{0.33\hsize}{!}{\includegraphics[angle=0]{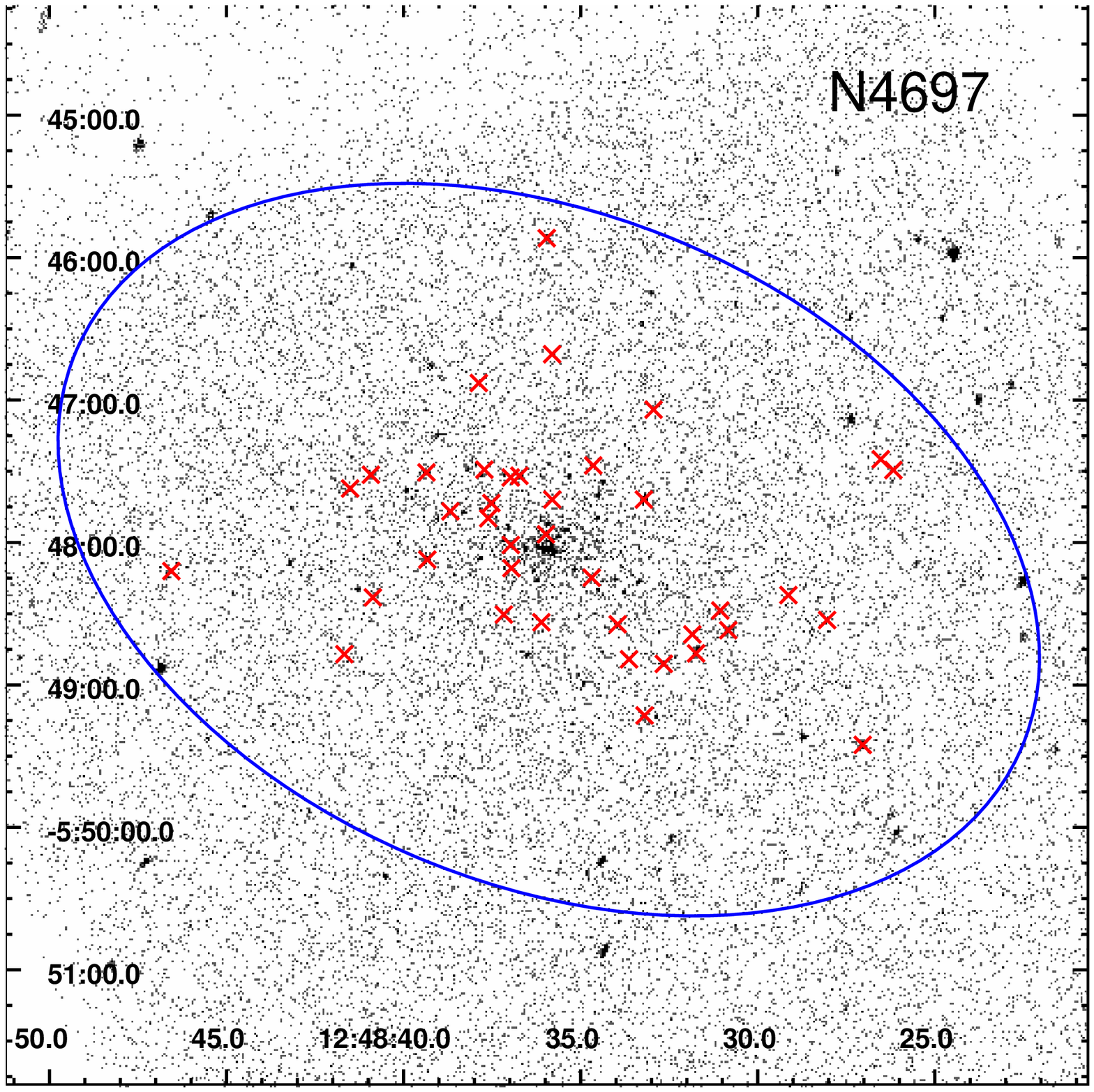}}
\resizebox{0.33\hsize}{!}{\includegraphics[angle=0]{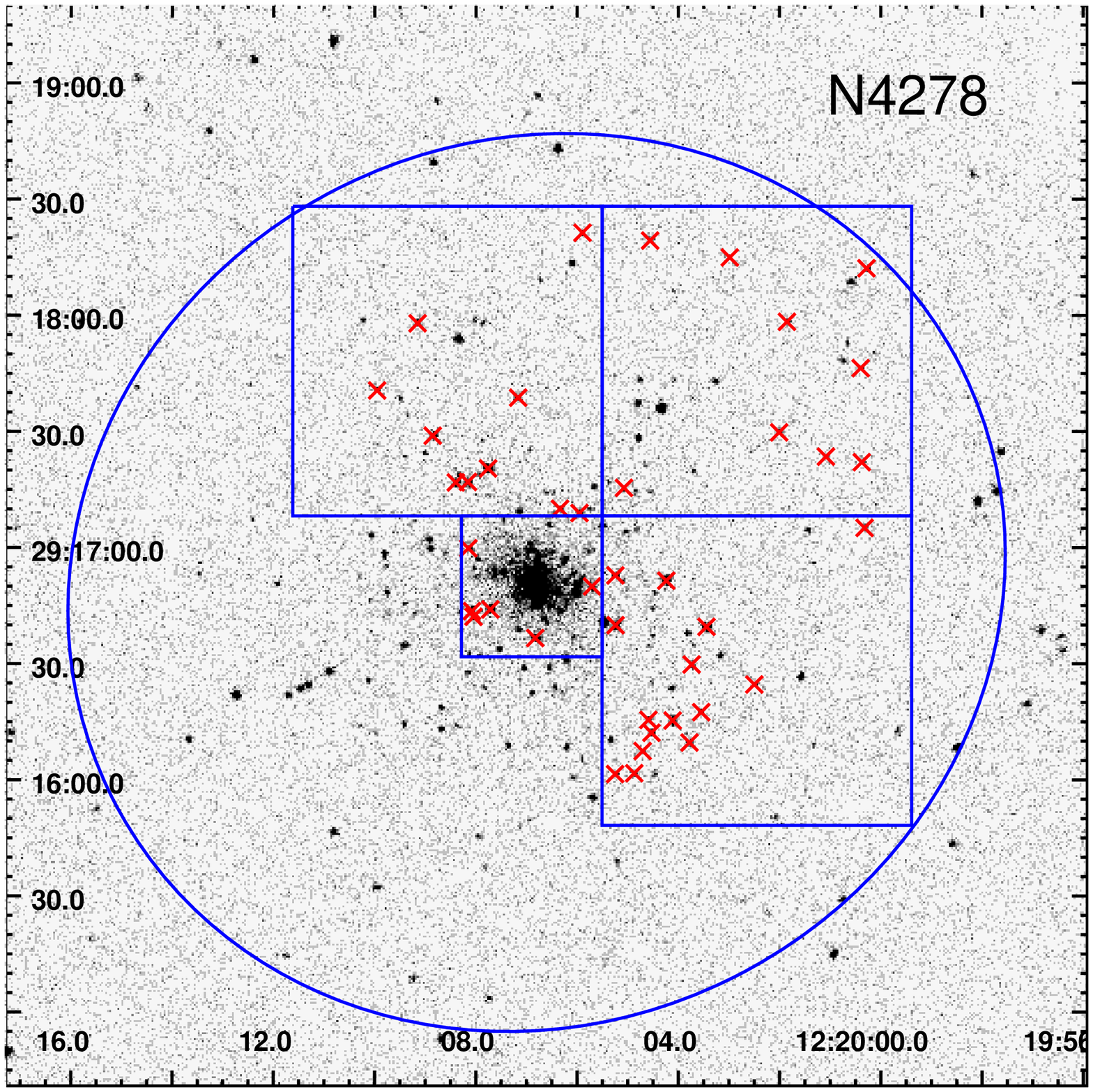}}
\caption{X-ray images (0.5--8 keV) of external galaxies and the fields of view studied. For M31 and M81 we study the sources within the ellipses with radii and position angle shown in Table \ref{tab:sample}. For Maffei 1, NGC 3379, NGC 4697, and NGC 4278 ellipses show D25 regions and squares show the HST fovs. The field of view of interest are the overlapping regions between the two. Crosses show detected GC-LMXBs in each galaxy. The $Chandra$ image of Centaurus A can be found in Fig. 1 in \citet{Voss2009}.}
\label{fig:galaxy}
\end{center}
\end{figure*}

\section{The sample}
\label{sec:sample}

In constructing the sample our goal is to provide uniform coverage over as wide a range in luminosity as possible. We aim to study sources as faint as $10^{35}$ erg/s. On the other hand, our goal is to have good enough statistics at the bright end where the specific frequencies of sources (per GC or per unit stellar mass) is low. Therefore our strategy is to include all galaxies with the best sensitivities achieved by $Chandra$ so far and complement this with several sufficiently massive galaxies with somewhat low sensitivity in order to properly sample the high-luminosity domain.
We based our selection on the list of normal galaxies available in the public $Chandra$ archive. We did not exclude late-type galaxies, but in constructing the XLF of the field sources we considered only their bulges (to exclude possible contamination by HMXBs).
The main selection criterion used was a detection sensitivity better than
$\log(L_X)\sim 36.5-37$. This translates into a joint constraint on the
distance to the galaxy and the exposure time of the $Chandra$ observation. We also decided to exclude galaxies with stellar mass less than $10^{10}M_\odot$ because of their smaller LMXB populations and the consequently higher contamination by resolved CXB sources. Finally, we required the availability of extensive GC data in order to reliably separate GC and field sources. Our final sample includes seven nearby 
galaxies (Table \ref{tab:sample}). In addition we also include the GC sources in the Milky Way.

M31 is the only nearby galaxy with $Chandra$ sensitivity better
than 10$^{35}$ erg s$^{-1}$. \citet{Voss2007a} analysed 160 ksec of $Chandra$ data available at the time and found 12 LMXBs in confirmed GCs in the bulge. Since this study, an additional $\sim 140$ ksec of data has been collected by $Chandra$ (which brings the total exposure time of the bulge to over 300 ksec), and more accurate GC data have been published \citep{Peacock2010}. We also analysed an additional 160 ks observation of a region in the disk. 
Centaurus A was the target of a recent $Chandra$ VLP program. With a total $Chandra$ exposure time of $\sim$800 ks, 
a detection sensitivity of 6$\times$10$^{35}$ erg s$^{-1}$ has been reached in this galaxy.
\citet{Voss2009} find 47 GC-LMXBs in this galaxy, so we use their source lists in our analysis. 
A similar detection sensitivity was reached in M81 with an exposure time of $\sim$240 ks. The four other external galaxies in our sample have detection sensitivities of a few $\times$ $10^{36}$ erg s$^{-1}$ and are included to increase the statistics of luminous sources.
One of them, Maffei 1, is relatively small and marginally passed our mass threshold. However, it appears to be particularly rich in X-ray sources. The  
X-ray populations in NGC 3379, NGC 4697 and NGC 4278 have been studied previously  \citep[e.g.,][]{Kundu2007,Kim2009,Brassington2009}. For these galaxies we redid the data analysis and found it to be in overall agreement with the above authors. 

The Milky Way hosts 150 GCs \citep{Harris1996}, of which 12 are known to host bright LMXBs. As all the Milky Way GCs have been surveyed in the X-rays multiple times by various instruments, we assume that our sample of GC-LMXBs is complete. We used the data from the All-Sky Monitor aboard RXTE to measure the luminosities of these sources. The advantages and shortcomings of such an approach are discussed in section \ref{sec:MW}.

\section{Data analysis}
\label{sec:analysis}

\subsection{Source detection}
\label{sec:galaxy}

\begin{figure}
\resizebox{8.7cm}{!}{\includegraphics[angle=270]{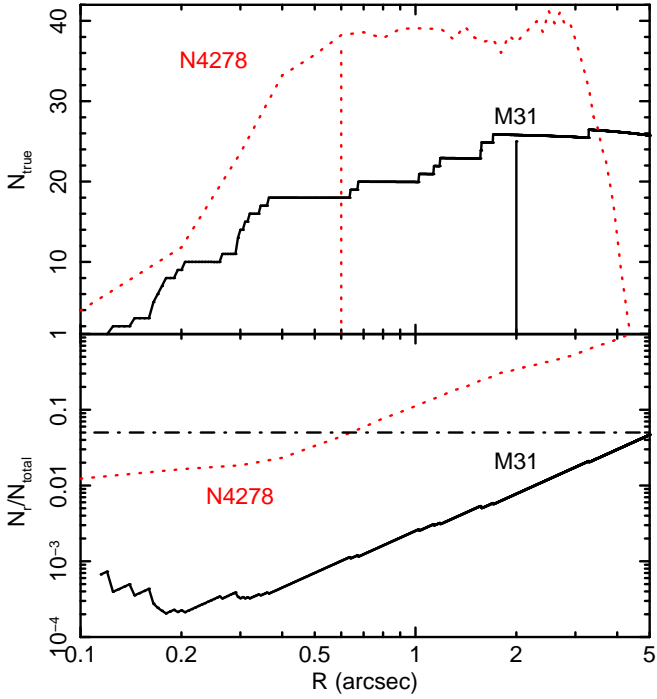}}
\caption{Examples of the determination of the optimal search radius for cross-correlation of X-ray source lists with GC catalogs for two galaxies -- M31 (solid lines) and NGC 4278 (dashed lines). 
Upper panel: the number of true matches ($N_{\rm true}$, computed as the difference between the number of total matches $N_{\rm total}$ the number of expected random matches $N_{\rm r}$) as a function of the search radius $R$. 
Lower panel: the ratio of the number of random matches to the number of total matches. The dash-dotted line is the 5\% level.
The vertical lines in the upper panel show our choice of the search radii for these two galaxies.}
\label{fig:ranmatch}
\end{figure}

The details of the $Chandra$ observations of external galaxies used here are listed in Table \ref{tab:observation}. These observations were reduced  following the standard CIAO threads (CIAO version 3.4; CALDB version 3.4.1). 
Time intervals when background flares happen were not excluded, since the benefit of the increased exposure time outweighs the increased background.
The energy range was limited to 0.5-8.0 keV. We have only used sources located in the regions where $Chandra$ data overlaps with the GC data to construct the XLFs. The sizes of these "study fields"  are listed in Table \ref{tab:sample}, and the regions are overlaid on the X-ray images shown in  Fig. \ref{fig:galaxy}. 

In order to detect sources we used ``wavdetect"  with the same parameters as \citet{Voss2006,Voss2007a}. Thresholds were set to be $10^{-6}$ for all 
galaxies, yielding an average of 1 false source per $10^6$ 0.492$\times$0.492 
arcsec$^2$ ACIS pixels. In order to prevent distortion of point sources due to attitude reconstruction errors in the course of  
combining images, we used the brightest sources detected within a 4$^{\prime}$ radius of the
telescope axis in each observation and corrected offsets between observations using 
the method described in \citet{Voss2007a}. For the disk region of M31, the
observations are distributed in too wide an area to make this correction possible. 
For M81 and NGC 4697, the offsets between observations were insignificant thus this step was skipped. For NGC 3379, we used offsets from \citet{Brassington2008}. 
The observations were then shifted to match the coordinate system of the reference (marked with an asterisk in Table \ref{tab:observation}) and thereafter combined together and re-analyzed.

The next step is to apply an absolute astrometry correction to the combined image. We used the 
2MASS All-Sky Point Source catalog \citep{Skrutskie2006} and correlated it with 
the brightest X-ray point sources. X-ray images
were shifted to give the shortest rms-distances between the X-ray sources and their 
2MASS counterparts. These corrections are listed in Table \ref{tab:sample}, where dx refers to correction in the west and dy is the correction to the north. 
In the case of three galaxies: NGC 3379, NGC 4697, and NGC 4278, this step was skipped because we did not find enough matches. 

The statistics of the detected sources are summarized in Tables \ref{tab:source1} and \ref{tab:source2}. 
To estimate the count rates we followed the method described in \citet{Voss2007a}. 
The luminosities of point sources were calculated assuming a power-law spectrum with
$\Gamma$=1.7. Count rates in 0.5-8.0 keV band were converted into absorption corrected fluxes 
assuming Galactic absorption. We list the conversion factors in Table \ref{tab:sample}. 

To estimate the expected numbers of CXB sources among detected X-ray sources, we used the results 
of the CXB log($N$)-log($S$) determination by \citet{Moretti2003}. We used their equation (2) for the soft band counts and 
converted the flux to the 0.5--8.0 keV band, assuming a power-law spectrum with a photon 
index of 1.4. The predicted numbers of CXB sources account for the incompleteness are described 
in section \ref{sec:icf}. The computed numbers of CXB sources are listed in Table \ref{tab:source1}. In the closest 
galaxies, background AGN account for a large fraction of detected X-ray sources, especially in M31 where 
nearly half of the X-ray sources are CXBs. Maffei 1 is a small galaxy that is abundant in LMXBs, and the contamination by CXB sources is minimal. In NGC 3379 and NGC 4278 HST the WFPC2 field-of-views (FOVs) are 
located in the very central region, where the CXB fraction is less than 10\%. In NGC 4697 the CXB fraction is 
about 15\% in the D25 region. 

The CXB estimates based on the average source counts are subject to uncertainties caused by angular fluctuations of the density of background AGN. These are likely to be reduced in our analysis as it covers a rather large solid angle composed of non contiguous fields. Nevertheless, for each individual galaxy we verified that the observed density of compact sources outside its main body is consistent, within the statistical errors, with the predicted density of CXB sources. This was possible to do directly for Maffei 1, NGC 3379,  NGC 4697, and NGC 4278, thanks to their relatively small angular size. For M31 and Centaurus A whose angular extent exceeds or is comparable to the Chandra FOV,  we used the results of \citet{Voss2007b} and \citet{Voss2009}. In Centaurus A the CXB source density was found to exceed the average source count by a factor of $\sim 1.5$, which that was accounted for in our calculations.

\begin{table*}
\begin{center}
\caption{Statistics of compact sources I}
\label{tab:source1}
\renewcommand{\arraystretch}{1.1}
\begin{tabular}{lrrrcrccc}
\hline
\hline
Galaxy        & $N_{\rm XRS}$ & $N_{\rm CXB}$ & $N_{\rm GC}$ & $K^{GC}_{\rm opt}$  & $N_{\rm GC-X}$       & $K^{GC,X}_{\rm opt}$ & $R$ & $N_{\rm r}$  \\
(1)           & (2)           &(3)            & (4)          & (5)                 & (6)                  & (7)                  & (8) & (9)          \\
\hline
M31           & 386           & 194           & 121          & 1.00                & 26                   & 1.00                 & 2.0$^{\prime\prime}$  & 0.2 \\
Cen A         & 231           &  64           & 479          & $0.67\pm0.03$       & 47                   & $0.55\pm0.19$        & 2.0$^{\prime\prime}$  & 1.2 \\
M81           & 220           &  79           &  77          & $0.77\pm0.09$       &  8                   & $0.84\pm0.48$        & 3.0$^{\prime\prime}$  & 0.8 \\
Maffei 1      &  38           &   1           &  20          & 1.00                &  4                   & 1.00                 & 1.0$^{\prime\prime}$  & 0.3 \\
N3379         &  59           &   4           &  61          & $0.80\pm0.10$       &  9                   & $0.74\pm0.39$        & 1.0$^{\prime\prime}$  & 0.6 \\
N4697         & 117           &  17           & 441          & $0.85\pm0.04$       & 39                   & $0.93\pm0.31$        & 0.8$^{\prime\prime}$  & 1.1 \\
N4278         & 120           &   6           & 266          & $0.69\pm0.04$       & 40                   & $0.86\pm0.29$        & 0.6$^{\prime\prime}$  & 1.8 \\
MW            & --            & --            & 150          & 1.00                & 12                   & 1.00                 &	      --             & --  \\
\hline
Total         & 1171          & 365           & 1615         & --                  & 185                  & --                   & --                    & 6.7 \\
\hline
\end{tabular}
\end{center}
Columns are: (1) -- Galaxy name. (2) -- Total number of resolved X-ray point sources in the study fields. (3) -- Predicted number of CXB sources in the study fields above the corresponding sensitivity threshold. (4) -- Number of optically identified GCs. (5) -- Completeness fraction of GC lists and its 1$\sigma$ uncertainty, estimated as described in section \ref{sec:k_opt} (6) -- Number of LMXBs found in GCs. (7) -- Completeness fraction and its uncertainty of GC lists with respect to GCs containing  LMXBs (see section \ref{sec:k_opt}). (8) -- Search radius to match XRS to GC. (9) -- Expected number of random coincidences of X-ray sources with GCs.
\end{table*}

\begin{table*}
\begin{center}
\caption{Statistics of compact sources II}
\label{tab:source2}
\renewcommand{\arraystretch}{1.1}
\begin{tabular}{lcccrcccr}
\hline
\hline
Galaxy        & $N_{\rm GC-X}^{\rm 1}$ & $N_{\rm GC-X}^{\rm 2}$ & $N_{\rm GC-X}^{\rm 3}$ & $N_{\rm GC-X}^{\rm 3}/N_{\rm GC}$ & $N_{\rm F-X}^{\rm 1}$ & $N_{\rm F-X}^{\rm 2}$ & $N_{\rm F-X}^{\rm 3}$ & $N_{\rm F-X}^{3}/M_{*}$ \\
(1)           & (2) & (3) & (4) & (5) & (6) & (7) & (8) & (9) \\
\hline
M31           & 2 & 11 & 12 & $0.10\pm0.03$	& 110 & 64  & 28 & $9.5\pm1.8$  \\
Cen A         & 0 & 16 & 30 & $0.06\pm0.01$	& 6   & 85  & 29 & $7.2\pm1.3$  \\
M81           & 0 & 4  & 4  & $0.05\pm0.03$	& --  & --  & -- & --	        \\
Maffei 1      & 0 & 2  & 2  & $0.10\pm0.07$	& 0   & 6   & 12 & $14.8\pm4.3$ \\
N3379         & 0 & 0  & 8  & $0.13\pm0.05$	& 0   & 9   & 24 & $6.3\pm1.3$  \\ 
N4697         & 0 & 2  & 34 & $0.08\pm0.01$	& 0   & 2   & 66 & $8.9\pm1.1$  \\ 
N4278         & 0 & 4  & 36 & $0.14\pm0.02$	& 0   & 3   & 52 & $17.4\pm2.4$ \\
MW            & 1 & 9  & 2  & $0.013\pm0.009$	& --  & --  & -- &  --          \\
\hline
Total         & 3 & 48 & 128& $0.08\pm0.01$	& 116 & 169 & 211& $9.8\pm0.7$  \\
\hline
\end{tabular}
\end{center}
Columns are: (1) -- Galaxy name. Columns (2)--(4) and (6)--(8) -- Number of GC-LMXBs ($N_{\rm GC-X}$) and field LMXBs ($N_{\rm F-X}$) in different luminosity ranges (1, 2 and 3 refer to $\log(L_X)$ ranges of 35--36, 36--37 and $>$37) with incompleteness higher than 0.5. The source numbers are not corrected for incompleteness, and the CXB contribution is not subtracted. Columns (5) and (9) -- The specific number of GC-LMXB (per GC) and field LMXBs (per 10$^{10}$ $M_{\odot}$) in the highest luminosity bin $\log(L_X)>37$. The numbers are corrected for incompleteness of X-ray source lists, the contribution of CXB is subtracted. Note that the specific numbers of GC-LMXB are not corrected for incompleteness of the GC lists and are given here as a characterization of our sample, rather than of the properties of GC systems in different galaxies.).
\end{table*}

\subsection{Identification of GC sources}
\label{sec:correlation}

We correlated the lists of detected X-ray point sources with the GC lists 
available for M31, M81, Maffei 1, NGC 3379, NGC 4697, and NGC 4278. For Centaurus A 
we used the results of \citet{Voss2009}.

For M31, the most recent and complete GC catalog is a systematic survey using WFCAM on the United Kingdom Infrared Telescope 
and SDSS by \citet{Peacock2010}. In total there are 416 confirmed GCs, with 121 located 
in the two $Chandra$ fields in our study. GCs in M81 are from 
\citet{Perelmuter1995}, \citet{Chandar2001}, and \citet{Schroder2002}. 
Chandar's study is based on deep HST observations that cover 25\% of our 
$Chandra$ field. We took the 59 GCs from \citet{Chandar2001} and the others 
from the other two catalogs, which resulted in 77 confirmed GCs in
the $Chandra$ field. For Maffei 1, there are 20 GCs from HST observation by \citet{Buta2003}. 
For NGC 3379, we took the 61 GCs from \citet{Kundu2001}, which are based on deep HST 
observations. The GC list for NGC 4697 is taken from \citet{Jordan2011}. And the GCs in 
NGC 4278 are from \citet{Kundu2001} and \citet{Brassington2009}.

The search radius $R$ used in cross-correlating the X-ray source lists and GC catalogs was 
chosen for each galaxy individually based on the following considerations.  
The number of random matches is
$N_{\rm r}=\pi R^{2}\times N_{\rm XRS} \times N_{\rm GC}/A$, 
where $N_{\rm XRS}$ is the number of X-ray point sources, $N_{\rm GC}$ the number of 
GCs, and $A$ the area of our study field. Because of the rather high source density, the number of random matches may be non-negligible for high values of the search radius $R$. On the other hand, the search radius has to be broad enough to account for position errors and, for the closest galaxies, the finite angular sizes of GCs.   
We therefore devised a procedure in which we varied  $R$ from 0 to $5^{\prime\prime}$ (Fig. \ref{fig:ranmatch}). For each value of $R$ we computed the number of true matches as the number of total detected matches minus the predicted number of false matches calculated from the above formula. This number increases with $R$ and saturates at some value of $R$ that depends on the typical positional error and angular extent of GCs. This value of $R$ may be chosen as the optimal match radius. In some cases, however, it results in too high a fraction of false matches in the sample. We therefore set an additional requirement that the predicted number of false matches does not exceed $5\%$ of the total number of matches. This procedure is a simplified version of the method used in \citet{Shtykovskiy2005}. The optimal search radius used for the program galaxies are listed in Table \ref{tab:source1}. As expected, there is a general trend that nearby galaxies require larger search radii. 

The numbers of X-ray sources associated with GCs are listed in Table \ref{tab:source1} along with the predicted numbers of false matches.

\subsection{Incompleteness of GC lists}
\label{sec:k_opt}

Although the availability of the high quality GC optical data was one of the criteria in selecting our galaxy sample, the GC lists are not 100\% complete for all of them. The incompleteness of these lists can result in incompleteness of the GC-LMXB lists and can compromise the shape and (less importantly) the normalization of the GC XLF.    

\begin{figure}
\resizebox{\hsize}{!}{\includegraphics[angle=270]{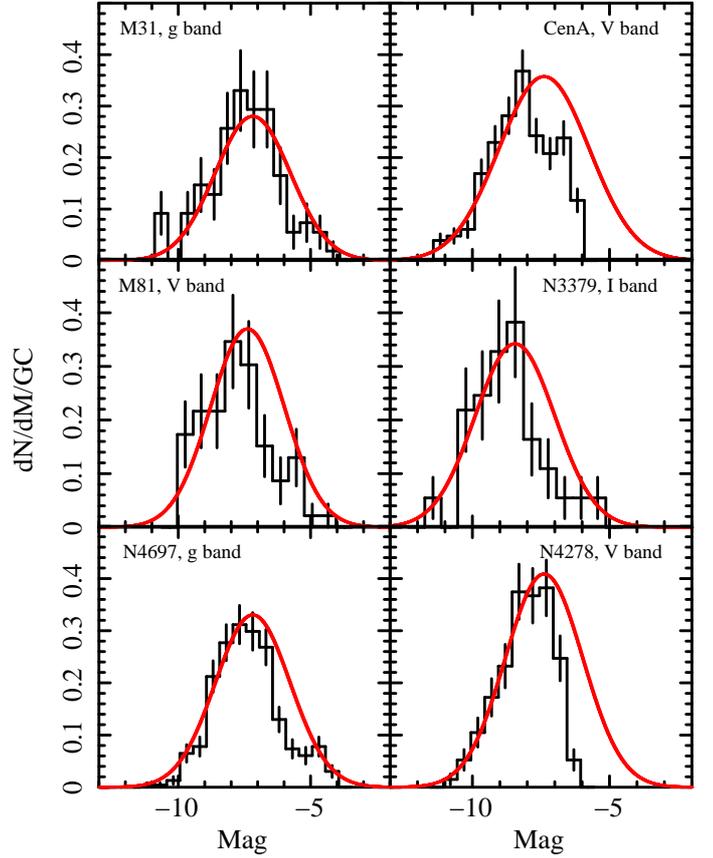}}
\caption{The observed luminosity functions of GCs for six galaxies in our sample and their best-fit models. The turnover luminosity of the model in different bands was fixed at the values determined elsewhere (see text), the width $\sigma$ was a free parameter of the fit.}
\label{fig:gclf}
\end{figure}

In order to estimate the completeness fraction of the GC lists we used the fact that optical luminosity function of GCs (GCLF) can be described to good accuracy by log-normal distribution in the form
\begin{equation}
\frac{dN}{dM} = \frac{1}{\sqrt{2\pi}\,\sigma}
                       \exp\left[-\frac{(M-\mu)^2}{2\sigma^2}\right]\ ,
\label{eq:gauss}
\end{equation}
where $M$ is the absolute magnitude of GC, $\mu$ the turnover luminosity, and $\sigma$ the dispersion. The turnover luminosity is remarkably constant in different galaxies.  We used the following values for different bands:  $\mu^{0}_{V}=-7.41$, $\mu^{0}_{I}=-8.46$ \citep{Kundu2001} and $\mu^{0}_{g}=-7.2$ \citep{Jordan2007a}.  The reddening corrected photometry data of  GCs for  each galaxy (see references in section \ref{sec:correlation}) was fit by this model using maximum likelihood method. The fitting was performed using only GCs above the completeness limit of the optical data for each galaxy. The width of the distribution $\sigma$ and normalizations were free parameters of the fit.  The data along with the  best-fit model are shown for six galaxies from our sample in Fig. \ref{fig:gclf}. The completeness fraction of the GC lists $K^{GC}_{opt}$ was then determined as a ratio of the total number of detected GCs (of all magnitudes)  to the total number predicted by integrating the best-fit model. 
The results are listed in  Table \ref{tab:source1}. Given the completeness limits of M31 and Maffei 1 data, their GC lists are complete. The list of GCs in the  Milky Way is also believed to be reasonably complete \citep{Harris2001}. In agreement with this, the best-fit values obtained for these three galaxies are consistent, within errors, with 1. We therefore set the completeness fraction for these three galaxies equal to unity. 

\begin{figure}
\resizebox{\hsize}{!}{\includegraphics[angle=0]{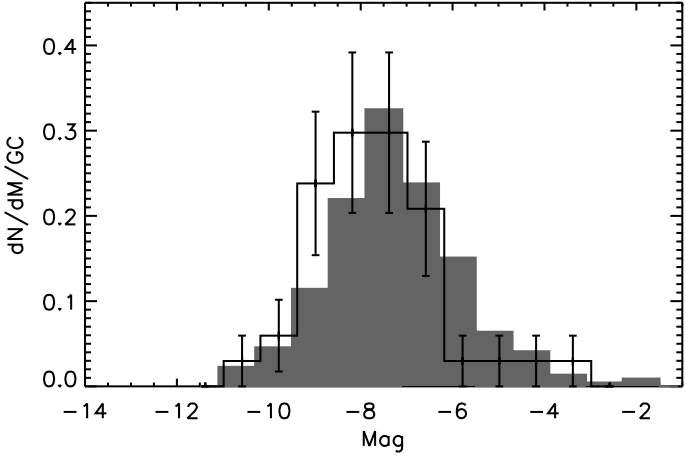}}
\caption{The combined LFs of all GCs in the Milky Way, M31, and Maffei 1 (the shaded histogram). The thin solid histogram shows the combined LF of GCs hosting X-ray sources. The Poisson errors for the latter are indicated by  the vertical error bars. The statistical errors are $\sim$2-3 times smaller for the combined LF of all GCs.}
\label{fig:gcgcxb}
\end{figure}

The second factor, required to correctly computing GC XLF is $K^{GC,X}_{opt}$ --  the  completeness fraction of the GC-LMXB identifications, caused by the incompleteness of the overall GC lists. If the probability of finding an LMXB in a globular was independent of its optical luminosity, the two quantities would coincide: $K^{GC,X}_{opt}=K^{GC}_{opt}$ would hold. However, it has been shown that X-ray sources tend to be associated with  brighter GCs \citep[e.g.,][]{Sivakoff2007}. This is illustrated by the Fig. \ref{fig:gcgcxb} where we plot the combined LF of all GCs and all the GCs hosting an X-ray source in the three galaxies which GC lists are complete -- MW, M31 and Maffei 1. To determine  $K^{GC,X}_{opt}$ we assume that LF of GCs hosting an X-ray source is the same in all galaxies. Using the combined LFs in Fig.\ref{fig:gcgcxb} as the template, we then use the ratio of  the numbers of GCs hosting X-ray sources above and below the threshold magnitude of $V=-7$ ($V=-8$ for Cen A) to estimate the number of missed X-ray sources in GCs in each galaxy. The threshold  magnitude  was chosen so that the GC lists are complete above its value. The results of this calculation, along with their uncertainties are listed in Table \ref{tab:source1}.

\subsection{The Milky Way sources}
\label{sec:MW}

The luminosities of the Milky Way sources were calculated from the light curves measured by the All-Sky Monitor aboard RXTE. The light curves were averaged over the period from January 1996 to June 2009. The count rates were transformed to the 0.5-8.0 keV band fluxes with the conversion factor obtained using PIMMS: 1 count s$^{-1}$ = 4.3$\times$10$^{-10}$ erg cm$^{-2}$ s$^{-1}$. A power-law  with $\Gamma$=1.7 was assumed. To compute source luminosities we used distances to the GCs from \citet{Harris1996}.  

The following comments regarding the determination of the luminosities of the Milky Way sources are in order. The ASM fluxes are averaged over a significantly longer time scale than the $Chandra$ data for external galaxies. Although both the ASM ($\sim$ years) and $Chandra$ ($\sim 1-10$ days of total integration time, $\sim$ years time span of observations)  integration time scales are much longer than the characteristic time scales of the accretion disk in these sources the variability of the X-ray light curves could in principle result in ``clipping" of the XLF i.e. smoothing out the extrema of long term variability. The effect of averaging ASM light curves on the XLF was studied by \citet{Postnov2005}, who came to the conclusion that flux probability distribution functions for persistent galactic X-ray binaries are such that light curve averaging does not modify the shape of the power-law luminosity distribution. To verify this further we considered variations in the XLF obtained by averaging ASM light curves over shorter intervals, comparable to the duration of $Chandra$ observations of external galaxies. The results of this analysis are presented in section \ref{sec:discussion}. However, the effect of such time averaging may be more significant for transients that, for long averaging times, will ``accumulate" in the low-luminosity bins and will lead to a steepening of the XLF  \citep{Voss2007a}.  It is also an issue for M31, which was observed in several short observations distributed over the time span of a few years, and to a lesser extent for M81. This can potentially lead to significant distortions of the XLF, depending on the average time and light curve properties of transients. However, as discussed in section \ref{sec:caveats}, it is not a significant factor in our particular case.

It is also known that one of the Galactic GCs, M15, contains two bright X-ray binaries \citep{White2001}. Transients have been detected in two other GCs,  NGC 6440 \citep{Heinke2010} and Terzan 5 \citep{Bordas2010}. The net effect of source blending on the XLF of GCs was considered and shown to be negligible in \citet{Voss2007a}. It is also discussed in section \ref{sec:discussion}.


\begin{figure}
\resizebox{\hsize}{!}{\includegraphics[angle=270]{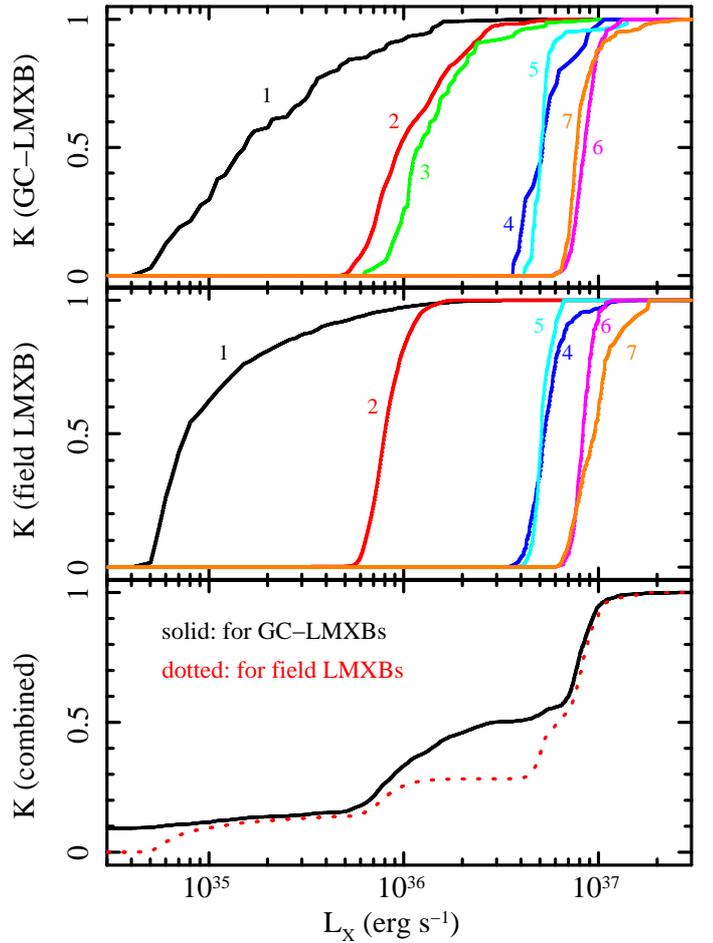}}
\caption{The incompleteness functions for individual galaxies from our sample (top panel for GC-LMXBs and middle panel for field LMXBs;  the numbers 1, 2, 3, 4, 5, 6, and 7 by the curves refer to M31, Centaurus A, M81, Maffei 1, NGC 3379, NGC 4697, and NGC 4278, correspondingly). The bottom panel shows the combined curves for GC (solid line)  and field (dotted line) sources. L$_X$ is the unabsorbed X-ray luminosity in 0.5-8 keV.}
\label{fig:icf}
\end{figure}

\section{Combined XLFs of GC and field LMXBs}
\label{sec:combined_xlf}

In combining the data from individual galaxies two effects need to be considered: correction for the incompleteness and removal of the contamination by background AGN. The former has to compensate for the fact that different sensitivities have been achieved for different galaxies, as well as for variations in the source detection sensitivity across the FOV in individual observations. On the other hand, estimation of the contribution of CXB sources in the XLF has to take into account the effects of incompleteness that affect the detection of CXB sources as well.  

There are a few different weighting schemes for combining the XLFs; here, we have used the one that produces the best signal-to-noise ratio under the assumption that XLFs of different galaxies have the same shape (Eqs. (\ref{eq:xlf1}) and (\ref{eq:xlf2}) below). Obviously this assumption can only be verified to the accuracy allowed by the statistical quality of the individual XLFs, which is by a factor of a few less than the accuracy of the combined XLF. Indirectly, this assumption is supported by not seeing large variations in the specific frequency of GC and field sources between galaxies (Table \ref{tab:source2}, but see also section \ref{sec:caveats}).

\subsection{X-ray incompleteness correction}
\label{sec:icf}

We calculated the X-ray incompleteness function $K_X(L)$ following the method of \citet{Voss2006}. For each pixel in the study area the point source detection sensitivity was calculated by inverting the detection method and using the actual local PSF and background \citep{Voss2007a}. $K_X(L)$ is computed as the fraction of pixels, weighted by the assumed spatial distribution of sources, in which a source with the given or higher luminosity would be detected. This computed quantity is the detection efficiency as a function of source luminosity.
Because it depends on the spatial distribution of sources, we compute it separately for different source populations. 
For CXB sources we assume a flat spatial distribution. Field LMXBs are assumed to follow the $K$-band light, as reported in the 2MASS Large Galaxy Atlas \citep{Jarrett2003}. 
For GC-LMXBs no weighting was applied, but only pixels containing GCs were used in the calculation.  
The incompleteness function for GC and field sources in individual galaxies are shown in the two upper panels in  Fig.\ref{fig:icf}.

The bottom panel in Fig.\ref{fig:icf} shows the combined incompleteness functions for GC and field sources computed by  summing individual incompleteness functions weighted by the number of GC and stellar mass inside the study area of each galaxy. Since the distributions of GCs and stellar mass do not differ strongly, one may expect these two functions to be nearly identical. This is in fact the case throughout most of the luminosity range. The two curves diverge near $\sim 5\times 10^{36}$  erg s$^{-1}$ because of the different areas used to study GC and field sources in Centaurus A \citep[see][for details]{Voss2009}. The difference below $\sim 10^{35}$ erg s$^{-1}$ is caused by only using GC sources for the Milky Way.

\subsection{LF of GC-LMXBs}
\label{sec:gc_xlf}

There are 185 GC-LMXBs in total in our sample. To avoid uncertainties from the highly incomplete low-luminosity end, we adopted a completeness threshold of 0.5 and used the curves shown in the upper panel of Fig.\ref{fig:icf} for each galaxy to determine the corresponding luminosity limit. Sources below these limits were excluded from the XLF construction. This procedure excluded source numbers 13, 39, 98, 107, 108, and 109 from Table \ref{tab:longtable}. 
The XLF value in the j-th luminosity bin centered at $L_j$ and having a width of $\Delta \log(L_j)$  was computed according to the equation:
\begin{equation}
\left( \frac{dN}{d\log(L_j)}\right)^{GC} = 
\frac{1}{\Delta{\log(L_j)}}
\sum_{k=1}^{N_{gal}} \sum_{L_i^k\in \Delta L_j} \frac{1}{N^{GC}_{eff}(L_i^k)}\ ,
\label{eq:xlf1}
\end{equation}
where $N^{GC}_{eff}$ is the effective number of GCs involved in the calculation of the XLF at the given luminosity $L_X$, corrected for optical and X-ray incompleteness. It depends on X-ray luminosity because it accounts for the X-ray incompleteness, 
\begin{equation}
N^{GC}_{eff}(L)=\sum_{k=1}^{N_{gal}}N^{GC}_k \  \frac{K^{GC,X}_{opt}}{K^{GC}_{opt}} \  K^{GC}_{X,k}(L) \ ,
\label{eq:ktot2}
\end{equation}
where $N^{GC}_k$ is the number of observed GCs in the study field of the $k-$th galaxy, $K^{GC}_{X,k}(L)$ is the X-ray incompleteness function for GC sources in the $k-$th galaxy, the $K^{GC,X}_{opt,k}$ and  $K^{GC}_{opt,k}$ are optical completeness factors, described in section \ref{sec:k_opt} and listed in Table \ref{tab:source1}. The thus computed XLF is normalized per GC. 

The effective X-ray incompleteness of the GC data can be defined as 
\begin{equation}
K^{GC}_{tot}(L)=\frac{\sum_{k=1}^{N_{gal}}N^{GC}_k K^{GC}_k(L)}{\sum_{k=1}^{N_{gal}}N^{GC}_k}\ . 
\label{eq:ktot1}
\end{equation}
This is the quantity plotted in Fig.\ref{fig:icf} (individual $K(L)$ not clipped at the incompleteness level of 0.5 when plotting the figure).

The factor $K^{GC,X}_{opt}/K^{GC}_{opt}$ in eq.(\ref{eq:ktot2}) accounts for the incompleteness of the optical GC data.  The  denominator in this expression, $K^{GC,X}_{opt}$, is rather poorly constrained by our data (Table \ref{tab:source1}) and is consistent with  $K^{GC,X}_{opt}=K^{GC}_{opt}$ within the measurement uncertainties. In fact, given the amplitude of the uncertainties, using the best-fit values would introduce additional noise into the obtained XLF. We therefore assumed that  $K^{GC,X}_{opt}=K^{GC}_{opt}$ in the further calculations. With this assumption the incompleteness of the optical data cancels out (see the discussion in the beginning of the last paragraph in section \ref{sec:k_opt}). The impact of this assumption on the final LF of GC-LMXBs is investigated in section \ref{sec:caveats}. 

In the case of GC sources, the contamination by background AGN is insignificant so was ignored.
Indeed, the predicted total number of random matches between resolved CXB sources and GC positions with the given  search radii from Table \ref{tab:source1} is $\approx 1.1$. The final XLF of GC sources is shown in Fig.\ref{fig:xlf_GC_field}. The incompleteness-corrected number of GC sources with luminosity exceeding 10$^{35}$ erg s$^{-1}$ is $\approx 244$.

\subsection{LF of field LMXBs}
\label{sec:field_xlf}

We have only considered field sources in elliptical galaxies and the bulges of spiral galaxies, in order to minimize the 
contamination by HMXBs. M81 was not included due to the relatively small size of its bulge. Thus, the bulge regions of M31, Maffei 1, NGC 3379, NGC 4278, and NGC 4697 were combined with the Centaurus A observations. For the last, we excluded the jet and radio lobe regions that have small-scale structures in the diffuse emission \citep{Voss2009}. 
In M31 we excluded the sources located in the central $1^{\prime}$ of the galaxy since they have been demonstrated as very likely 
dynamically formed \citep{Voss2007a}. A separate XLF was constructed for these 36 sources, as discussed in the next section.  In Maffei 1, NGC 3379, NGC 4278, and NGC 4697 the central 
10$^{\prime\prime}$ were excluded. These regions are affected by source confusion, which makes accurate luminosity estimates difficult. We then followed a procedure similar to the GC sources by applying a luminosity threshold corresponding to $K(L)=0.5$ in each galaxy. With these selection criteria we obtained 496 sources above 
$10^{35}$ erg s$^{-1}$, of which  $\sim$177 are predicted to be CXB sources. 

\begin{figure}
\resizebox{\hsize}{!}{\includegraphics[angle=0]{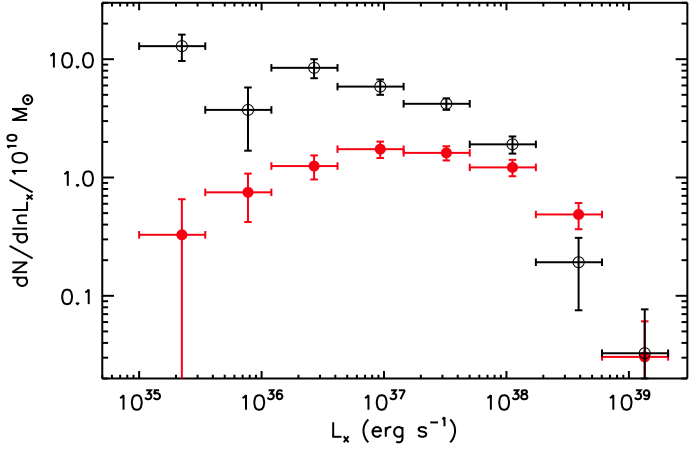}}
\caption{The combined XLFs of LMXBs in GC (filled circles) and in the field (open circles). L$_X$ is the unabsorbed X-ray luminosity in 0.5-8 keV. The contribution of CXB sources was subtracted and the incompleteness correction was applied. The field XLF is normalized to stellar mass of $10^{10}~M_\odot$. The GC XLF is normalized to have the same number of sources as the field XLF above $10^{38}$ erg s$^{-1}$. }
\label{fig:xlf_GC_field}
\end{figure}

In order to correctly subtract CXB contribution one has to take into account the difference in the incompleteness functions for CXB sources and LMXBs:
\begin{eqnarray}
\left( \frac{dN}{d\log(L_j)}\right)^{LMXB} = 
\frac{1}{\Delta{\log(L_j)}}    
\sum_{k=1}^{N_{gal}}\left( \sum_{L_i^k\in \Delta L_j} \frac{1}{K^{ LMXB}_{tot}(L_i^k)} \right. \nonumber \\ 
\left. - \int_{L\in \Delta L_j} 4\pi D_k^2\ \frac{dN_{CXB}}{dL} \frac{K^{\,CXB}_k(L)}{K^{LMXB}_{tot}(L)} dL
\right)\ ,
\label{eq:xlf2}
\end{eqnarray}
where $D_k$ is the distance to the $k-$th galaxy, $K_{tot}^{LMXB}(L)$ the combined incompleteness function for LMXBs computed similar to eq.(\ref{eq:ktot1}), and $dN^{\,CXB}/dS$ the $\log(N)-\log(S)$ distribution for the CXB sources.
In practice we implemented this by adding a large number ($\sim$$10^3$) of fake sources with small negative weights to each galaxy's source list. This accounted for the CXB $\log(N)-\log(S)$ distribution and incompleteness function of the galaxy. The sum of these weights for each galaxy equals the predicted number of CXB sources in this galaxy. These ``enhanced" source lists were used to produce the combined XLF according to eq.(\ref{eq:xlf1}): 
\begin{eqnarray}
\left( \frac{dN}{d\log(L_j)}\right)^{LMXB} = 
\frac{1}{\Delta{\log(L_j)}}
\sum_{k=1}^{N_{gal}}\left( \sum_{L_i^k\in \Delta L_j} \frac{1}{K^{LMXB}_{tot}(L_i^k)}
\right. \nonumber \\
\left. - \sum_{L^{cxb}_i\in \Delta L_j} \frac{w^{cxb}_i}{K^{LMXB}_{tot}(L_i^{cxb})}
\right)\ .
\label{eq:xlf2}
\end{eqnarray}
The final XLF for field sources normalized to unit stellar mass is shown in Fig.\ref{fig:xlf_GC_field}. The total stellar mass involved in this calculation is 1.82$\times$10$^{11}M_\odot$. The specific frequency of LMXBs above $10^{36}$($10^{37}$) erg s$^{-1}$ is 25.7(9.6) per 10$^{10}M_\odot$, 
which is consistent with the average values from \citet{Gilfanov2004} -- 33.9(14.3).

\begin{table}
\begin{center}
\caption{Comparison of XLFs of different populations.}
\label{tab:comparison}
\renewcommand{\arraystretch}{1.3}
\begin{tabular}{ccc}
\hline
\hline		      
        & $L_{min}=10^{35}$ & $L_{min}=10^{36}$  \\
\hline
$R_{GC}$                  & $0.89^{+0.18}_{-0.16}$ & $0.71^{+0.13}_{-0.12}$  \\
$R_{F}$                   & $3.82^{+0.65}_{-0.61}$ & $1.67^{+0.28}_{-0.27}$ \\ 
$R_{C}$                   & $2.18^{+1.07}_{-0.64}$ & $1.91^{+0.97}_{-0.58}$ \\       
$P (R_{F}<R_{GC})$    & $<$$10^{-7}~~(>5\sigma)$  & $3.6\cdot10^{-4}~~(3.6\sigma)$  \\
$P (R_{C}<R_{GC})$    & $8.3\cdot 10^{-3} ~~(2.6\sigma)$ & $4.6\cdot10^{-3}~~(2.8\sigma)$  \\
\hline
\end{tabular}
\end{center}
Note -- $R_{GC}$, $R_{F}$, and $R_{C}$ are ratios of the number of faint sources ($L_{min}<L_x<10^{37}$ erg s$^{-1}$) to the number of bright sources ($L_x>10^{37}$ erg s$^{-1}$) for GCs, field sources, and sources in the inner $1\arcmin$ of M31. $P$ is the probability that the luminosity distributions of corresponding populations are drawn from the same mean (see text for details).
\end{table}


\section{Results}
\label{sec:result}

\begin{figure}
\resizebox{\hsize}{!}{\includegraphics[angle=0]{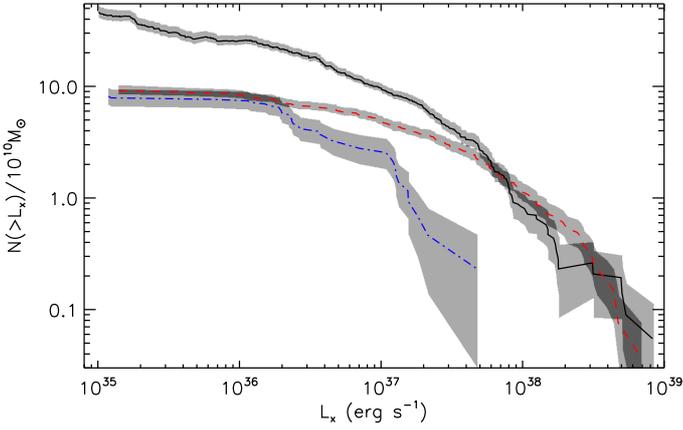}}
\caption{The combined XLFs of LMXBs in different environments plotted in the cumulative form. L$_X$ is the unabsorbed X-ray luminosity in 0.5-8 keV. The contribution of CXB sources was subtracted and the incompleteness correction was applied. The field XLF (solid) is normalized to the stellar mass of $10^{10}~M_\odot$. The normalizations of GC (dashed) and M31 nucleus (dash-dotted) XLFs are arbitrary. The shaded areas around the curves show $1\sigma$ statistical uncertainty.
}
\label{fig:xlf_cum}
\end{figure}

The background-subtracted and incompleteness-corrected XLFs of the GC and field LMXBs are shown in Figs. \ref{fig:xlf_GC_field} and \ref{fig:xlf_cum}. The XLF of field sources is normalized to a stellar mass of $10^{10}$ M$_\odot$. The XLF of GC sources is normalized to the same number of sources above $10^{38}$ erg/s as the field XLF. It is obvious from the plot that the two luminosity distributions have different shapes. Although they differ across the entire luminosity range,  the most evident difference is at lower luminosities, below $\log(L)\sim 37$. 
Both XLFs change their slope between $\log(L)\sim 37-38$. Due to their rather complicated shapes we did not attempt to do global fits with analytical functions. Instead, we perform power-law fits to the high- and low-luminosity ends. We did maximum-likelihood fits to the background-subtracted XLFs. To account for the incompleteness, we multiplied the model by the respective incompleteness function. In the $\log(L)\ge 38$ range, we obtained differential slopes of  $1.70^{+0.60}_{-0.58}$ and $2.06^{+0.92}_{-0.75}$ for GC and field sources correspondingly. At the faint end, $\log(L)\le 37$ the slopes are: $0.68^{+0.21}_{-0.23}$ and $1.17^{+0.13}_{-0.14}$ respectively. The slopes of the field sources are broadly consistent with the parameters of average LMXB XLFs from \citet{Gilfanov2004}.

Differences in the incompleteness curves and in the CXB contribution render direct application of the K-S test to compare these two XLFs impossible. We have therefore considered the ratio of the number of faint to bright sources in order to assess the statistical significance of the difference between the two XLFs, the same method as used in \citet{Voss2009}.
For each population we computed the ratio $R=N_{\rm faint}/N_{\rm bright}$, with the boundary between faint and bright set to $10^{37}$ erg s$^{-1}$. We ran Monte-Carlo simulations to calculate statistical errors and the significance of our results. The details of these calculations are described in \citet{Voss2009}. For each XLF we did $10^{7}$ Monte-Carlo runs. The results are listed in Table \ref{tab:comparison}. The lines marked ``P'' give the probability of obtaining the observed values of R due to statistical fluctuations, while their mean (true) values obey the relation given in parenthesis. These numbers can be interpreted as the probability that the corresponding luminosity distributions are drawn from the same mean. The upper limit of $<10^{-7}$ in the left column means that no such realizations were detected in $10^7$ Monte-Carlo runs. 

These calculations show that the GC and field XLFs differ at a confidence level of $<10^{-7}$, which corresponds to a significance of $>5\sigma$. To investigate the robustness of this conclusion we have also used a more restricted luminosity range of $\log(L_X)>36$, where the incompleteness functions vary less and different galaxies from our sample make more uniform contributions. In this case the confidence level decreases to $3.4\cdot 10^{-4}$ ($3.6\sigma$), but the conclusion still holds.

\section{Discussion}
\label{sec:discussion}

Our analysis has revealed a significant difference in the luminosity distributions of field and GC sources. A similar result was previously reported for a few nearby galaxies \citep{Voss2007a,Voss2009,Kim2009}, some of which are included in our sample. However, limited numbers of sources in these galaxies (even in Centaurus A) did not permit any detailed investigation of the luminosity distributions, but merely suggested a deficit of faint sources among dynamically formed  LMXBs. Given  the sensitivity expected in a typical $Chandra$ observation and distances to the closest massive galaxies,  the main limitation of these studies -- insufficient number of sources cannot be lifted when investigating a single galaxy. This motivated us to combine data for several nearby galaxies. Thus, we have produced average XLFs of different populations of LMXBs with far better statistical quality than achieved  before.  With these XLFs we confirm the general conclusion of previous studies that the fraction of faint sources in GCs is a factor of $\sim 4$ smaller than sources in the field. Moreover, the overall behavior is more complex and cannot be described in terms of a simple roll-over of the XLF at low luminosities. Instead, luminosity distributions of GC and field sources differ throughout the entire luminosity range. 

It has long been debated whether the entire population of LMXBs in galaxies, including those in the field, was formed dynamically in GCs  \citep{white2002,Kundu2002,Kundu2007,Irwin2005,Juett2005,Humphrey2008}. In this scenario it is suggested that either field LMXBs are systems expelled from their GCs, or the host cluster itself was destroyed leaving the LMXB to join the field population. The significantly different luminosity distributions of LMXBs residing in GCs and in the field argues against this scenario. As is well known, the mass transfer rate in a Roche lobe-filling system is defined by its orbital period, mass ratio, and the evolutionary stage of the donor star. The distributions of these parameters are obviously different in the population of primordially formed binaries and in the dynamically formed systems in GCs. It is therefore not surprising that these populations have different mass transfer rates distributions and, correspondingly, different XLFs.\footnote{As a caveat, we mention that the cross-section for the dynamical interaction and, therefore, the probability for a binary to be ejected from a GC, is a strong function of the orbital period of the binary. Therefore, in the GC-scenario one may also expect different orbital period  distribution of the binaries retained in GCs and those  ejected into the field. This may lead to different XLFs of the two sub populations, if ejection of binaries is the main mechanism of populating the field LMXBs, rather than GC destruction. Detailed populations synthesis calculations are required in order to see if this may explain the observed XLFs.}

\begin{figure}
\resizebox{\hsize}{!}{\includegraphics[angle=0]{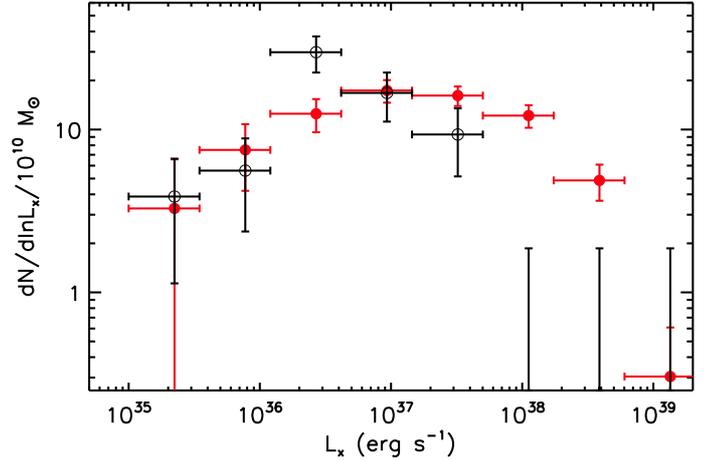}}
\caption{The XLFs of sources in the inner 1$^{\prime}$ of M31(open circles) and  of GC sources (filled circles). L$_X$ is the unabsorbed X-ray luminosity in 0.5-8 keV. The normalization of the GC XLF is arbitrary. No sources are detected in the three highest luminosity bins of the M31 XLF.}
\label{fig:xlf_dynamical}
\end{figure}

\subsection{Sources in the nucleus of M31}

\citet{Voss2007a} demonstrate that most sources in the inner $\sim 1\arcmin$ of M31 are very likely to have formed dynamically, similar to the sources in GCs. In particular, they find that their spatial distribution follows the $\rho^2_*$ law, in contrast to the X-ray sources outside this region, where the density is proportional to the stellar density.
We updated the LMXB list in this region using an exposure approximately twice of the one presented in \citet{Voss2007a}. We have detected three new sources, bringing the total number to 36 (The increase in the number of sources for a $\log(N)-\log(S)$ distribution with the slope of $-1$ would be $\sim$13). We excluded one source that coincided with a GC and computed the luminosity distribution of the detected sources, performing incompleteness correction and CXB subtraction as described before.  The resulting XLF  is shown in Fig. \ref{fig:xlf_dynamical} along with the XLF of GC sources. It is obvious from the plot that the two distributions have similar shapes at $\log(L_X)\la 37$ but differ at the bright end, with the XLF of the sources in the M31 nucleus having a deficit of bright sources.
To test the statistical significance of this conclusion we ran the same tests as we did to compare XLFs of GC and field LMXBs in Sec \ref{sec:result}. We found that the LF of LMXBs in the M31 nucleus differs from the GC XLF with a significance $\approx$2.6-2.8$\sigma$ (Table \ref{tab:comparison}).
In other words, one should expect 26 sources with $\log(L_X)>37$ in the nucleus of M31, assuming that both distributions have the same shape and using the number of faint sources for normalization. The observed number of bright sources is 11, which is $\approx 3\sigma$ less than expected. Both calculations give similar results, confirming the marginal significance of our conclusion.

Although both populations were formed dynamically, there is an important difference between stellar environments in GCs and galactic nuclei: stellar velocities in the latter are about $\sim 10-20$ times higher. This leads to different formation channels in GCs and galactic nuclei \citep[and references therein]{Voss2007b}. Calculations of \citet{Voss2007b} suggest that in the high-velocity environment of the M31 nucleus the main formation channel for X-ray binaries may be tidal captures of compact objects by low-mass stars, producing short orbital period binaries. In GCs, in contrast, LMXBs are predominantly formed in exchange reactions and collisions of neutron stars with evolved stars. Obviously, this difference will affect the distributions of binary systems over the mass accretion rate. Detailed population synthesis calculations are required to understand what this effect may be and to interpret the observed luminosity distributions in a more quantitative and meaningful manner.

\begin{figure}
\resizebox{\hsize}{!}{\includegraphics[angle=0]{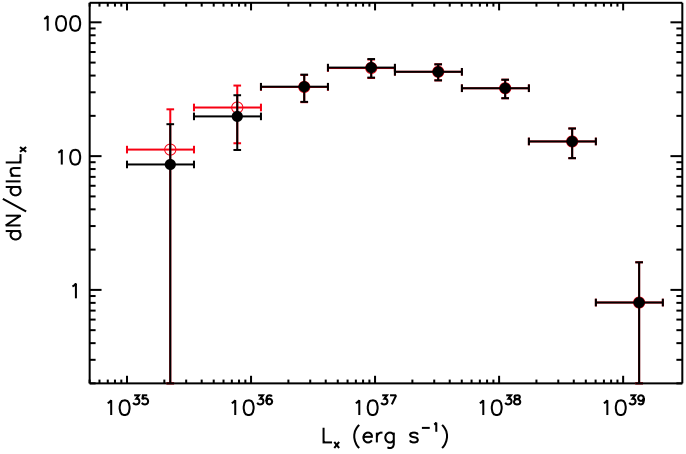}}
\caption{The maximum possible effect of the optical incompleteness on the GC LMXB XLF. The solid symbols show the XLF computed assuming $K^{GC,X}_{opt}=K^{GC}_{opt}$ (our default version), open circles -- assuming $K^{GC,X}_{opt}=1$ (the maximum possible correction). To emphasize the effect on the shape, rather than overall normalization, the XLF are normalized to the total number of detected GCs. See sections \ref{sec:k_opt},  \ref{sec:gc_xlf}. and \ref{sec:caveats} for details.}
\label{fig:k_optx}
\end{figure}

\subsection{Caveats}
\label{sec:caveats}

Several caveats regarding possible systematic effects are in order. The distance uncertainties, in the limit of several galaxies, will smooth out the luminosity distributions. For a smaller sample (which is the case for the current study), there could be a non trivial effect on the computed XLFs. However, the distances to the galaxies in our sample are fairly well known, with an accuracy of $\sim 5-15\%$ (Table \ref{tab:sample}). This translates to $\sim 10-30\%$ uncertainty in the luminosity and $\sim 4-11\%$ in its logarithm. This is a factor of $\sim 4-12$ smaller that the bin width used in the XLF calculations. Thus it should not affect the measured XLFs in any significant way.

We used a rather inhomogeneous set of the GC lists, having varying degree of completeness. The procedure of correction for incompleteness of the optical data is described in section \ref{sec:k_opt} and \ref{sec:gc_xlf}. Its accurate implementation, however, was hindered by the large statistical uncertainties of the  completeness fraction of the GCs hosting an X-ray source,  $K^{\,GC,X}_{opt}$. As  results of section \ref{sec:k_opt} were consistent with $K^{GC,X}_{opt}=K^{GC}_{opt}$ (the latter much better constrained), we assumed that this relation holds for all galaxies.  This could be the case, for example, if the probability of finding an LMXB did not depend on the optical luminosity of the GC. However, this is known not to be the case \citep[e.g.,][]{Sivakoff2007} (see Fig. \ref{fig:gcgcxb}). As is obvious from Eq.(\ref{eq:ktot2}),  the optical incompleteness would have the strongest effect on the XLF if for all galaxies $K^{GC,X}_{opt}=1$ (i.e. if all LMXBs were located in the brightest GCs and not subject to optical incompleteness at all, which is not true either). To illustrate its amplitude, we show in Fig.\ref{fig:k_optx} the XLFs computed in these two limiting cases. As is obvious from the plot, the XLF does not change by more than  $\sim 20-30\%$ in the two lowest luminosity bins. We emphasize that the example shown in the plot illustrates  the maximum possible effect of the optical incompleteness, the real effect being smaller.

Combining XLFs necessarily involves an assumption regarding the similarity of their shapes in individual galaxies. Although we did not detect statistically significant differences between different galaxies, this assumption cannot be verified directly at the same level of accuracy as provided by the output average XLF. On the other hand, we do detect marginally significant variations in the specific frequency of X-ray sources in GCs between galaxies, although these may be related, at least in part, to the incompleteness of the GC lists in more distant galaxies. However, if they are real, they may be accompanied by variations in the XLF shapes.
The effect of such variations may be further amplified by the fact that data for different luminosities come from different galaxies. The low-luminosity domain, $\log(L)<36$, is covered exclusively by the nearby M31 and Milky Way, whereas the bright end is dominated  by sources located in more massive but more distant galaxies, such as Centaurus A and NGC 4697. This is another unavoidable limitation, as bright sources, although more easy to detect, are less frequent, and it takes a bigger galaxy to have them in large numbers. On the other hand, bigger galaxies are more distant and the sensitivity achieved in a typical $Chandra$ observation is lower. Conversely the nearby galaxies, where fainter sources can be studied, tend to be less massive and contain fewer bright sources. Luckily, the 800 ksec Very Large $Chandra$ program on Centaurus A and relatively good coverage of M31 allowed us to bridge faint, intermediate, and bright luminosity ranges.

\begin{figure}
\resizebox{\hsize}{!}{\includegraphics[angle=0]{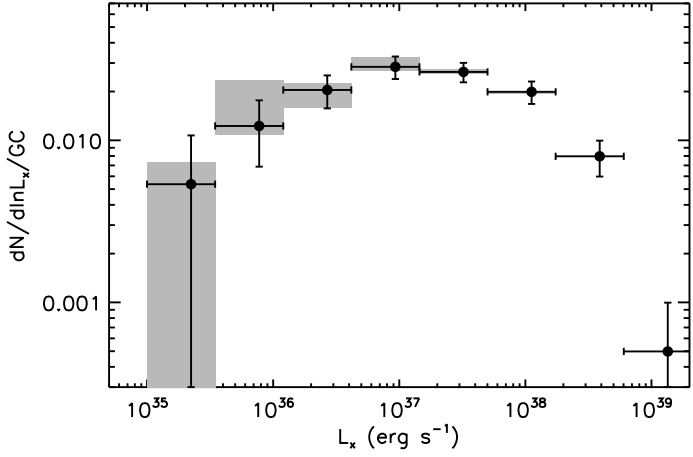}}
\caption{The combined XLF of GC-LMXBs. L$_X$ is the unabsorbed X-ray luminosity in 0.5-8 keV. The XLF uncertainty shown by the shaded regions is due to the variability of GC-LMXBs 
in the Milky Way.}
\label{fig:xlferror_MW}
\end{figure}

About half of the GCs with X-ray sensitivity in the lowest luminosity domain, $\log(L)\la36$ are located in the Milky Way. The flux determination of the latter, based on the averaging of the ASM light curves, may be subject to systematic effects. It is different from those affecting $Chandra$ galaxies data as discussed in section \ref{sec:MW}.
Primarily, this is due to the long integration times of ASM light curves. To investigate its effect on the XLF, we divided the ASM light curves into 100 sub intervals with a duration of 50 days each (comparable to the integration time of the longest $Chandra$ observations) and recalculated the GC XLF 100 times, each time using the data from different sub intervals to compute ASM fluxes for the Milky Way sources. The range of obtained XLF values is shown by shaded area in Fig. \ref{fig:xlferror_MW}. As is obvious from the plot, the long integration time of ASM data does not affect the GC XLF significantly.

\begin{figure}
\resizebox{\hsize}{!}{\includegraphics[angle=0]{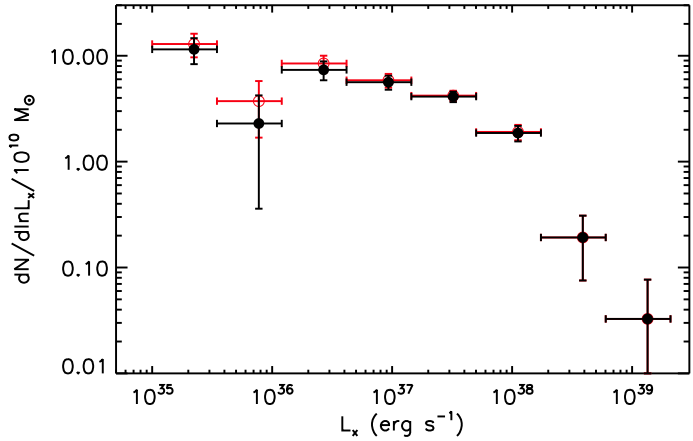}}
\caption{The combined XLFs of field sources with (open circles) and without (filled circles) transient sources in
M31. L$_X$ is the unabsorbed X-ray luminosity in 0.5-8 keV. Due to multiple short observations averaging the luminosities of transients  (which are bright in only a few observations and faint in many others) the number of faint sources is artificially increased, making the XLF appear steeper.  See text for details.}
\label{fig:xlferror_m31}
\end{figure}

If the time span of observations is longer than the typical time scale of transient sources, averaging of their luminosity can also modify the shape of the luminosity distribution, making it steeper  \citep{Voss2007a}. The typical decay time scales of the Galactic transient sources are in the  $\sim$weeks--months range. Thus, for the Milky Way GCs, this issue  is addressed by the above exercise with the ASM light curves, and Fig.\ref{fig:xlferror_MW} demonstrates that averaging of transients does not result in significant modifications of XLF, given its statistical quality. This issue is also relevant for the multiple Chandra observations of M31 and, to a less extent, M81.
Indeed, the  $Chandra$ image of  the bulge of M31 was obtained by combining  more than 40 short ($\sim$5 ks) observations. As transients are bright only in a few observations and dim in many others, they will tend to accumulate in the low-luminosity bins, making the XLF steeper.  In \citet{Voss2007a} 28 transients were reported, two of which  (Src. 22 and Src. 35 in Table \ref{tab:longtable}) are in our GC-LMXB source list and 21 are in the field source list.  We recomputed the luminosity distributions excluding these sources and did not find any significant changes (Fig. \ref{fig:xlferror_m31}). The results of the statistical tests reported in Table \ref{tab:comparison} are not changed significantly either: the probability $P (R_{F}<R_{GC})$ for the full luminosity range remains $<10^{-7}$ (although this may be affected somewhat by the incompleteness of the transient list at the faint end) for sources with $\log(L_X)>36$ it changes from $3.6\cdot 10^{-4}$ to $2\cdot10^{-3}$ (from $3.6\sigma$ to $3.1\sigma$). 
This proves that the contamination by transient sources in M31 does not significantly bias our results. It is much less significant for other galaxies, as they were observed by $Chandra$ in much fewer longer observations. 
Thus, averaging of persistent and transient sources does not lead to significant (as compared to statistical errors) distortions of the XLFs derived in this paper and does not affect our results in any significant way. This conclusion should not be taken out of the context though. In a more general case, the effects discussed above may be important and may need a more elaborate treatment.

Another factor that can modify the apparent XLF of GC LMXBs is the multiplicity of X-ray sources in GCs, which can affect both Milky Way data and $Chandra$ data for external galaxies. We use the Milky Way GCs to estimate its impact on XLF. One of the Milky Way GCs, M15, is known to contain two persistent LMXBs  (X2127+119-1 and X2127+119-2) with instantaneous  luminosities of 9.5$\times$10$^{35}$ erg s$^{-1}$ and 1.5$\times$10$^{36}$ erg s$^{-1}$ (converted to 0.5-8 keV band) \citep{White2001}. Obviously these two sources could not be resolved by ASM, which measured the long-term average luminosity of 4.05$\times$10$^{36}$ erg s$^{-1}$. Similarly, they would not be resolved by $Chandra$ in any of our external galaxies with the exception of M31, where it may be marginally possible. Two bright transients have been detected recently, in NGC 6440  \citep{Heinke2010} and Terzan 5 \citep{Bordas2010} with the luminosities in the $\sim 10^{36}-10^{38}$ erg/s range. Their effect on the "snapshot" XLF would depend on their unknown duty cycle. Assuming a duty cycle of $\sim 0.5$, which seems to be a very generous upper limit, the multiplicity fraction for the Milky Way GCs is $\sim 1/12-2/12\approx 8-16\%$. As demonstrated in \citet{Voss2007a}, the multiplicity at the level of $\sim 10\%$ does not modify the luminosity distribution in any significant way. We also checked to see how the GC XLF is affected if the ASM M15 source is replaced by two source with the luminosities determined by $Chandra$ and the transient source in NGC 6440 at its brightest state is added to the sample. The overall effect on the XLF is insignificant and the $R_{GC}$ changed from $0.89^{+0.18}_{-0.16}$ to  $0.92^{+0.17}_{-0.14}$, which is also negligible. We conclude that unless the multiplicity is much higher in external galaxies, it does not affect our conclusions in any significant way.

\section{Summary}
\label{sec:conclusion}

The aim of this study was to produce accurate luminosity distributions of LMXBs in different environments -- dynamically formed systems in GCs, in the nucleus of M31 and field sources of presumably primordial origin -- in order to facilitate their quantitative comparison and to provide input for verifying population synthesis models. This goal required a broad luminosity coverage with a point source detection sensitivity reaching  $10^{35}$ erg/s and, on the other hand, good sampling of the high-luminosity end, where the specific frequency of sources (per GC or per unit stellar mass) is low. As this combination of properties cannot be achieved with a single galaxy, we combined the data from a number of galaxies. To this end, we assembled a sample of galaxies from the public $Chandra$ archive which is best suited to our study. It included seven nearby galaxies (M31, Maffei 1, Centaurus A, M81, NGC 3379, NGC 4697, and NGC 4278) and the Milky Way. We detected 185 X-ray sources in 1615 GCs, 36 sources in the nucleus of M31, and 998 sources in the fields of galaxies  (of which $\approx 365$ are expected to be background AGN).  These sources were used to produce the average luminosity distributions of different populations. In doing so we took special care to accurately subtract resolved CXB sources and correct for incompleteness effects. As a result, we produced XLFs of LMXBs with a statistical accuracy that far exceeds what has been achieved in previous studies.  

We demonstrate that, although  the luminosity distributions of LMXBs in different environments are similar in a broad sense (e.g., when compared with XLF of HMXBs), their detailed shapes are different. Although the fraction of faint LMXBs ($\log(L_X)<37$) in GCs is $\sim 4$ times smaller than in the field, in agreement with a suggested effect found in previous studies, the difference in their XLFs cannot be described merely in terms of a roll over of the XLF of GC sources. Rather, the luminosity distributions of these two populations of LMXBs appear to be different throughout the entire luminosity range. This may present a challenge for the models suggesting that the entire LMXB population was formed dynamically  in GCs and then expelled to the field due to kicks, dynamical interactions, or GC destruction.

We also compare luminosity distributions of LMXBs in the nucleus of M31 (its inner $1\arcmin$) and in GCs. We find that although their shapes at the low-luminosity end are similar (and different from the field sources), the M31 nuclear population appears to have far fewer luminous sources than GCs (and field population). For example, the most luminous source in the nucleus of M31 has the luminosity of $4.7\times10^{37}$ erg/s. If the XLFs were drawn from the same parent distribution, we would expect to see 11 sources above this luminosity, whereas we found none. Different estimates of the statistical significance of the  difference between the two XLFs give results in the $\sim 2.5-3\sigma$ range. The difference between the XLFs is likely caused by the factor of $\sim 10-20$ difference in stellar velocities in GCs and galactic nuclei, which leads to different dynamical formation channels. However detailed population synthesis calculations are needed in order to understand the particular mechanisms responsible for forming the observed luminosity distributions.

\begin{acknowledgements} 
We thank Mark B. Peacock for the help with the GC catalog in M31 and {\'A}kos Bogd{\'a}n for helpful discussions. This research made use of $CHANDRA$  
archival data provided by $CHANDRA$ X-ray Center, 2MASS Large Galaxy Atlas 
data provided by NASA/IPAC infrared science archive, and ASM RXTE data from 
HEASARC online service. Andr{\'e}s Jord{\'a}n acknowledges support from 
BASAL CATA PFB-06, FONDAP CFA 15010003, MIDEPLAN ICM Nucleus P07-021-F 
and Anillo ACT-086. We thank the referee for helpful remarks on this paper. 

\end{acknowledgements} 


\bibliographystyle{aa}
\bibliography{ms}

{\tiny
\longtab{6}{

\begin{longtable}{llccrllccr}
\caption{All the 185 LMXBs in GCs in our sample. }\\
\hline
\hline
Number & Galaxy & RA(J2000) &  DEC(J2000)& Luminosity & Number & Galaxy & RA(J2000) & DEC(J2000) & Luminosity \\
(1)    & (2)    & (3)       &  (4)       & (5)        & (1)    & (2)    & (3)       & (4)        & (5)        \\   
\hline
\endfirsthead
\caption{continued.}\\
\hline
\hline
Number & Galaxy & RA(J2000) & DEC(J2000) & Luminosity & Number & Galaxy & RA(J2000) & DEC(J2000) & Luminosity \\
(1)    & (2)    & (3)       &  (4)       & (5)        & (1)    & (2)    & (3)       & (4)        & (5)        \\   
\hline
\endhead
\hline
\endfoot
\label{tab:longtable}

1 & MW     & +17:35:47.64 & -30:28:55.70 & 0.14   & 74  & Cen A & +13:26:00.72 & -43:09:40.32 & 46.48 \\  
2 & MW     & +18:53:04.89 & -08:42:19.70 & 1.14   & 75  & Cen A & +13:25:46.56 & -42:57:02.88 & 55.91 \\
3 & MW     & +17:48:53.54 & -20:22:02.00 & 1.31   & 76  & Cen A & +13:25:35.52 & -42:59:34.80 & 57.54 \\
4 & MW     & +18:35:44.00 & -32:58:55.40 & 1.68   & 77  & Cen A & +13:25:09.12 & -42:58:58.80 & 58.10 \\
5 & MW     & +17:50:45.54 & -31:17:32.50 & 1.76   & 78  & Cen A & +13:25:10.32 & -42:53:32.64 & 69.64 \\
6 & MW     & +17:48:55.73 & -24:53:40.10 & 1.78   & 79  & Cen A & +13:25:12.96 & -43:01:14.16 & 79.97 \\
7 & MW     & +05:14:06.59 & -40:02:37.00 & 2.34   & 80  & Cen A & +13:25:31.68 & -43:00:02.88 & 87.42 \\
8 & MW     & +21:29:58.33 & +12:10:02.80 & 4.05   & 81  & Cen A & +13:25:35.28 & -42:53:00.96 & 97.94 \\
9 & MW     & +17:33:24.06 & -33:23:16.20 & 5.00   & 82  & Cen A & +13:25:54.48 & -42:59:25.44 & 105.59 \\
10 & MW    & +17:27:33.25 & -30:48:07.40 & 6.79   & 83  & Cen A & +13:25:07.68 & -43:01:14.88& 199.09 \\
11 & MW    & +17:50:12.66 & -37:03:08.20 & 14.70  & 84  & Cen A & +13:25:02.64 & -43:02:43.08 & 255.53 \\
12 & MW    & +18:23:40.57 & -30:21:40.60 & 66.12  & 85  & Cen A & +13:25:42.00 & -43:10:41.52 & 315.01 \\
13 & M31   & +00:42:29.64 & +41:17:57.27 & 0.04   & 86  & M81   & +09:56:05.30 & +69:06:43.53 & 2.01  \\
14 & M31   & +00:42:50.86 & +41:10:33.72 & 0.41   & 87  & M81   & +09:55:37.26 & +69:02:07.57 & 2.36  \\
15 & M31   & +00:43:14.65 & +41:25:13.32 & 0.84   & 88  & M81   & +09:55:51.97 & +69:07:39.18 & 4.83  \\
16 & M31   & +00:42:27.43 & +40:59:35.63 & 1.05   & 89  & M81	& +09:55:22.05 & +69:05:18.93 & 6.68  \\
17 & M31   & +00:42:34.40 & +40:57:09.31 & 1.07   & 90  & M81	& +09:55:54.93 & +69:00:56.03 & 35.33 \\
18 & M31   & +00:43:15.48 & +41:11:25.69 & 1.17   & 91  & M81   & +09:55:47.00 & +69:05:51.09 & 67.26 \\
19 & M31   & +00:42:40.60 & +41:10:33.60 & 1.39   & 92  & M81   & +09:55:58.54 & +69:05:26.04 & 70.40 \\
20 & M31   & +00:42:25.04 & +40:57:18.78 & 2.09   & 93  & M81   & +09:55:49.80 & +69:05:31.93 & 434.56 \\
21 & M31   & +00:42:41.43 & +41:15:23.71 & 2.53   & 94  & Maffei 1 & +02:36:37.26 & +59:39:15.50 & 5.41 \\
22 & M31   & +00:42:47.81 & +41:11:13.66 & 2.56   & 95  & Maffei 1 & +02:36:30.84 & +59:39:34.70 & 9.11 \\
23 & M31   & +00:43:07.51 & +41:20:19.44 & 3.24   & 96  & Maffei 1 & +02:36:26.03 & +59:39:06.91 & 16.22 \\
24 & M31   & +00:42:33.10 & +41:03:29.86 & 4.23   & 97  & Maffei 1 & +02:36:36.50 & +59:38:42.03 & 42.07 \\
25 & M31   & +00:42:09.51 & +41:17:45.42 & 9.31   & 98  & N3379	 & +10:47:50.47 & +12:34:23.11 & 4.14 \\
26 & M31   & +00:42:26.05 & +41:19:14.81 & 9.94   & 99  & N3379	 & +10:47:51.57 & +12:35:36.01 & 14.18 \\   
27 & M31   & +00:42:12.17 & +41:17:58.62 & 11.51  & 100 & N3379	 & +10:47:54.20 & +12:35:29.49 & 39.83 \\   
28 & M31   & +00:43:03.31 & +41:21:21.60 & 12.02  & 101 & N3379	 & +10:47:50.47 & +12:34:36.94 & 53.20 \\ 
29 & M31   & +00:42:31.25 & +41:19:38.78 & 18.50  & 102 & N3379	 & +10:47:50.33 & +12:35:06.59 & 58.00 \\
30 & M31   & +00:43:02.93 & +41:15:22.47 & 22.50  & 103 & N3379	 & +10:47:51.08 & +12:35:49.25 & 87.35 \\
31 & M31   & +00:43:03.86 & +41:18:04.79 & 28.37  & 104 & N3379	 & +10:47:52.77 & +12:35:08.58 & 242.70 \\
32 & M31   & +00:42:59.86 & +41:16:05.64 & 33.75  & 105 & N3379	 & +10:47:50.19 & +12:34:55.34 & 333.36 \\
33 & M31   & +00:42:59.65 & +41:19:19.18 & 34.19  & 106 & N3379	 & +10:47:52.65 & +12:33:38.01 & 680.58 \\
34 & M31   & +00:42:18.64 & +41:14:01.74 & 36.23  & 107 & N4697	 & +12:48:32.94 & -05:47:04.04 & 7.23 \\
35 & M31   & +00:43:14.31 & +41:07:19.68 & 46.14  & 108 & N4697	 & +12:48:33.63 & -05:48:49.20 & 8.22 \\
36 & M31   & +00:43:37.29 & +41:14:43.63 & 47.94  & 109 & N4697	 & +12:48:29.13 & -05:48:22.15 & 8.26 \\
37 & M31   & +00:43:10.61 & +41:14:51.24 & 77.93  & 110 & N4697	 & +12:48:34.64 & -05:47:27.55 & 9.39 \\
38 & M31   & +00:42:15.84 & +41:01:14.32 & 123.26 & 111 & N4697	 & +12:48:26.52 & -05:47:24.91 & 9.53 \\
39 & Cen A & +13:25:41.76 & -42:57:00.00 & 0.90   & 112 & N4697	 & +12:48:35.80 & -05:47:41.90 & 10.13 \\
40 & Cen A & +13:25:11.04 & -43:01:31.80 & 1.79   & 113 & N4697	 & +12:48:37.60 & -05:47:49.79 & 10.26 \\
41 & Cen A & +13:25:29.28 & -42:57:46.80 & 1.90   & 114 & N4697	 & +12:48:34.68 & -05:48:14.82 & 11.08 \\
42 & Cen A & +13:25:14.88 & -43:00:48.96 & 1.92   & 115 & N4697	 & +12:48:37.16 & -05:48:30.34 & 11.76 \\
43 & Cen A & +13:25:42.00 & -43:03:19.44 & 2.15   & 116 & N4697  & +12:48:28.04 & -05:48:32.66 & 13.69 \\
44 & Cen A & +13:25:58.32 & -43:08:06.72 & 3.15   & 117 & N4697  & +12:48:40.86 & -05:48:23.12 & 14.88 \\
45 & Cen A & +13:25:43.20 & -42:58:37.20 & 3.57   & 118 & N4697  & +12:48:41.66 & -05:48:47.04 & 15.00 \\
46 & Cen A & +13:25:35.28 & -43:05:29.40 & 4.45   & 119 & N4697  & +12:48:37.71 & -05:47:29.32 & 16.78 \\
47 & Cen A & +13:24:49.20 & -43:05:12.12 & 4.81   & 120 & N4697  & +12:48:35.80 & -05:46:40.69 & 17.44 \\
48 & Cen A & +13:25:27.60 & -43:05:24.72 & 5.19   & 121 & N4697  & +12:48:31.84 & -05:48:38.70 & 23.44 \\
49 & Cen A & +13:25:32.40 & -43:04:40.44 & 6.25   & 122 & N4697  & +12:48:36.97 & -05:47:32.61 & 23.97 \\
50 & Cen A & +13:24:58.08 & -42:56:10.32 & 6.84   & 123 & N4697  & +12:48:33.95 & -05:48:34.46 & 25.58 \\
51 & Cen A & +13:25:14.16 & -43:02:42.72 & 7.01   & 124 & N4697  & +12:48:26.16 & -05:47:29.50 & 26.29 \\
52 & Cen A & +13:25:22.08 & -43:02:45.24 & 7.68   & 125 & N4697  & +12:48:36.95 & -05:48:10.80 & 30.36 \\
53 & Cen A & +13:25:30.24 & -42:59:34.80 & 7.83   & 126 & N4697  & +12:48:33.19 & -05:49:12.85 & 40.98 \\
54 & Cen A & +13:25:32.88 & -42:56:24.36 & 8.54   & 127 & N4697  & +12:48:37.87 & -05:46:52.81 & 42.12 \\
55 & Cen A & +13:24:50.40 & -43:04:50.88 & 9.16   & 128 & N4697  & +12:48:40.92 & -05:47:31.44 & 42.31 \\
56 & Cen A & +13:25:38.40 & -42:57:19.80 & 10.12  & 129 & N4697  & +12:48:31.05 & -05:48:28.66 & 46.15 \\
57 & Cen A & +13:25:12.00 & -42:57:12.96 & 10.27  & 130 & N4697  & +12:48:41.50 & -05:47:37.25 & 46.82 \\
58 & Cen A & +13:26:07.68 & -42:52:01.56 & 10.40  & 131 & N4697  & +12:48:36.10 & -05:48:33.61 & 60.54 \\
59 & Cen A & +13:26:05.28 & -42:56:32.64 & 10.71  & 132 & N4697  & +12:48:46.55 & -05:48:12.02 & 75.50 \\
60 & Cen A & +13:26:10.56 & -42:53:43.08 & 11.14  & 133 & N4697  & +12:48:38.67 & -05:47:46.88 & 91.24 \\
61 & Cen A & +13:25:05.76 & -43:10:30.36 & 11.57  & 134 & N4697  & +12:48:35.95 & -05:45:51.79 & 91.79 \\
62 & Cen A & +13:25:28.08 & -43:04:01.92 & 13.71  & 135 & N4697  & +12:48:31.73 & -05:48:46.73 & 97.43 \\
63 & Cen A & +13:25:03.12 & -42:56:24.72 & 13.84  & 136 & N4697  & +12:48:36.97 & -05:48:01.04 & 110.29 \\
64 & Cen A & +13:25:32.88 & -43:04:28.92 & 15.85  & 137 & N4697  & +12:48:32.65 & -05:48:51.11 & 125.21 \\
65 & Cen A & +13:25:52.80 & -43:05:46.32 & 20.97  & 138 & N4697  & +12:48:35.97 & -05:47:56.56 & 150.11 \\
66 & Cen A & +13:25:05.04 & -43:01:32.88 & 22.37  & 139 & N4697  & +12:48:39.35 & -05:47:30.48 & 168.46 \\
67 & Cen A & +13:25:39.84 & -43:05:01.68 & 23.62  & 140 & N4697  & +12:48:36.72 & -05:47:31.89 & 178.73 \\
68 & Cen A & +13:25:32.40 & -42:58:49.80 & 23.69  & 141 & N4697  & +12:48:37.51 & -05:47:43.40 & 192.56 \\
69 & Cen A & +13:25:18.48 & -43:01:15.96 & 24.05  & 142 & N4697  & +12:48:27.03 & -05:49:25.25 & 206.76 \\
70 & Cen A & +13:25:10.56 & -43:06:24.12 & 27.47  & 143 & N4697  & +12:48:30.83 & -05:48:36.93 & 308.63 \\
71 & Cen A & +13:26:19.68 & -43:03:19.08 & 30.51  & 144 & N4697  & +12:48:33.21 & -05:47:41.90 & 455.00 \\
72 & Cen A & +13:25:19.92 & -43:03:09.72 & 31.62  & 145 & N4697  & +12:48:39.32 & -05:48:07.22 & 474.28 \\
73 & Cen A & +13:25:10.32 & -42:55:09.48 & 37.62  & 146 & N4278	 & +12:20:04.55 & +29:18:19.33 & 7.97 \\

147 & N4278 & +12:20:00.39 & +29:17:46.37 & 8.53   & 167 & N4278  & +12:20:07.16 & +29:17:38.74 & 50.30 \\
148 & N4278 & +12:20:04.70 & +29:16:07.46 & 9.19   & 168 & N4278  & +12:20:04.53 & +29:16:12.19 & 50.37 \\
149 & N4278 & +12:20:02.98 & +29:18:14.97 & 9.78   & 169 & N4278  & +12:20:00.32 & +29:17:05.11 & 52.01 \\
150 & N4278 & +12:20:00.37 & +29:17:22.08 & 11.84  & 170 & N4278  & +12:20:03.77 & +29:16:09.66 & 58.85 \\ 
151 & N4278 & +12:20:05.24 & +29:16:01.51 & 12.03  & 171 & N4278  & +12:20:08.04 & +29:16:42.13 & 61.23 \\
152 & N4278 & +12:20:04.87 & +29:16:01.73 & 15.23  & 172 & N4278  & +12:20:09.15 & +29:17:57.95 & 66.65 \\
153 & N4278 & +12:20:03.54 & +29:16:17.50 & 17.76  & 173 & N4278  & +12:20:04.11 & +29:16:15.34 & 66.84 \\
154 & N4278 & +12:20:02.49 & +29:16:24.65 & 18.28  & 174 & N4278  & +12:20:08.07 & +29:16:43.61 & 71.32 \\
155 & N4278 & +12:20:05.89 & +29:18:21.35 & 18.72  & 175 & N4278  & +12:20:08.85 & +29:17:28.92 & 90.19 \\
156 & N4278 & +12:20:01.08 & +29:17:23.52 & 21.70  & 176 & N4278  & +12:20:08.39 & +29:17:16.85 & 113.49 \\
157 & N4278 & +12:20:04.59 & +29:16:15.51 & 22.01  & 177 & N4278  & +12:20:05.70 & +29:16:49.98 & 119.54 \\ 
158 & N4278 & +12:20:09.95 & +29:17:40.59 & 24.65  & 178 & N4278  & +12:20:08.15 & +29:17:16.97 & 125.36 \\
159 & N4278 & +12:20:05.07 & +29:17:15.46 & 26.87  & 179 & N4278  & +12:20:05.95 & +29:17:08.94 & 138.53 \\
160 & N4278 & +12:20:06.33 & +29:17:10.05 & 28.26  & 180 & N4278  & +12:20:07.71 & +29:16:44.05 & 144.98 \\
161 & N4278 & +12:20:08.14 & +29:16:59.83 & 28.32  & 181 & N4278  & +12:20:06.82 & +29:16:36.65 & 145.01 \\
162 & N4278 & +12:20:00.28 & +29:18:12.18 & 29.04  & 182 & N4278  & +12:20:05.24 & +29:16:39.97 & 265.89 \\
163 & N4278 & +12:20:01.85 & +29:17:58.35 & 30.38  & 183 & N4278  & +12:20:04.23 & +29:16:51.47 & 269.32 \\
164 & N4278 & +12:20:05.24 & +29:16:52.84 & 37.40  & 184 & N4278  & +12:20:03.44 & +29:16:39.55 & 292.63 \\
165 & N4278 & +12:20:02.00 & +29:17:29.78 & 46.63  & 185 & N4278  & +12:20:07.76 & +29:17:20.46 & 388.68 \\
166 & N4278 & +12:20:03.73 & +29:16:29.81 & 48.82  &     &        &              &              &        \\

\hline
\end{longtable}
(1) -- the sequence number.
(2) -- the galaxy where source is detected.
(3),(4) -- Right ascension and declination of source.
(5) -- X-ray luminosity in 0.5-8 keV, in unit of 10$^{36}$ erg s$^{-1}$. 
}}

\end{document}